\documentstyle[12pt,epsf]{article}
 
\input{epsf}
 
\newcommand{\beq}{\begin{equation}}
\newcommand{\eeq}{\end{equation}}
\newcommand{\barr}{\begin{eqnarray}}
\newcommand{\earr}{\end{eqnarray}}
\newcommand{\gf}{G_{\mbox{{\scriptsize F}}}}
\newcommand{\heff}{{\cal H}_{\mbox{{\scriptsize eff}}}}

\def\apar{A_{\|}}
\def\aperp{A_{\perp}}

\def\bep{\epsilon}
\def\bq{\begin{quote}}
\def\eq{\end{quote}}

\def\spose#1{\hbox to 0pt{#1\hss}}
\def\lsim{\mathrel{\spose{\lower 3pt\hbox{$\mathchar"218$}}
 \raise 2.0pt\hbox{$\mathchar"13C$}}}
\def\gsim{\mathrel{\spose{\lower 3pt\hbox{$\mathchar"218$}}
 \raise 2.0pt\hbox{$\mathchar"13E$}}}

\begin{document}
 
\begin{titlepage}
 
\begin{flushright}
CERN-TH/98-85\\
FERMILAB-PUB-98/093-T\\
IC/98/25\\
hep-ph/9804253
\end{flushright}

\vspace{0.3truecm}
\begin{center}
\boldmath
\large\bf
Extracting CKM Phases and $B_s$--$\overline{B}_s$ Mixing Parameters\\ 
\vspace{0.2truecm}
from Angular Distributions of Non-Leptonic $B$ Decays
\unboldmath
\end{center}
\vspace{0.3truecm}
\begin{center}
Amol S. Dighe\\[0.1cm]
{\sl The Abdus Salam International Centre for Theoretical Physics\\ 
34100 Trieste, Italy}\\[0.6cm]
Isard Dunietz\\[0.1cm]
{\sl Theoretical Physics Division, Fermi National Accelerator Laboratory\\
Batavia, IL 60510, USA}\\[0.6cm]
Robert Fleischer\\[0.1cm]
{\sl Theory Division, CERN, CH-1211 Geneva 23, Switzerland}
\end{center}
\vspace{0.15cm}
\begin{abstract}
\vspace{0.2cm}\noindent
Suggestions for efficiently determining the lifetimes and mass difference 
of the light and heavy $B_s$ mesons ($B_s^L, B_s^H$) from $B_s \to J/ \psi 
\phi, D_s^{*+}D_s^{*-}$ decays are given. Using appropriate weighting 
functions for the angular distributions of the decay products 
({\it moment analysis}), one can extract $(\Gamma_H, \Gamma_L, 
\Delta m)_{B_s}$.  Such a moment analysis allows the determination of 
the relative magnitudes and phases of the CP-odd and CP-even decay 
amplitudes. Efficient determinations of CP-violating effects occuring 
in $B_s \to J/\psi \phi, D_s^{*+} D_s^{*-}$ are discussed in the light of 
a possible width difference $(\Delta \Gamma)_{B_s}$, and the utility of 
this method for $B \to J/\psi K^*, D^{\ast +}_s \overline D^{\ast}$ 
decays is noted. Since our approach is very general, it can in principle be 
applied to all kinds of angular distributions and allows the determination of 
all relevant observables, including fundamental CKM 
(Cabibbo--Kobayashi--Maskawa) parameters, as well as tests of various aspects 
of the factorization hypothesis. Explicit angular distributions and 
weighting functions are given, and the general method that can be used 
for any angular distribution is indicated.
\end{abstract}
 
\vfill
\noindent
CERN-TH/98-85\\
April 1998
 
\end{titlepage}
 
\thispagestyle{empty}
\vbox{}
\newpage
 
\setcounter{page}{1} 

\section{Introduction}\label{intro}

Strategies for obtaining experimental insights into CP violation
and non-factorizable contributions to weak decays are of particular
interest for present particle physics.  The observables of angular 
distributions  can be
obtained in an efficient way by using an {\it angular
moment analysis} \cite{dqstl}--\cite{chinese}. In this approach, the  
observed
experimental data are weighted by judiciously chosen functions, which
project out any desired observable.
This strategy is an alternative to the usual {\it likelihood fit}
method \cite{fit}.  It is demonstrated that the moment analysis  
extracts all observables of measured angular distributions, such as  
the ones occurring in weak decays of pseudoscalars [$P \to V \ell  
\nu,~ X_J \ell \nu,~ VV,$ etc.].  This method is of general validity.
In our present paper, we apply the formalism to angular
distributions \cite{valencia,kp} of $B_s$ and $B$ meson decays into  
two vector-meson final states that are caused by $\bar b\to\bar s c \bar c$  
quark-level  transitions. By making use of the general formalism outlined in 
this paper, it is straightforward to derive weighting functions for 
other exclusive mesonic or baryonic transitions, governed for instance 
by $b \to c \overline  
u d,~ c \ell \overline \nu,~ u \ell \overline \nu,~c \to s \overline  
d u,~ s \ell^+ \nu,~d \ell^+ \nu$.  

The mixing between neutral $B_s$ mesons is expected to give
rise to CP-even ($B_s^L$) and CP-odd ($B_s^H$) mass eigenstates, 
which may
have a perceptible width difference $\Delta\Gamma\equiv\Gamma_H-
\Gamma_L$~\cite{deltagamma}. Using appropriate weighting functions
for the angular
distributions of the decay products in the transitions $B_s\to J/
\psi\phi$ and/or $B_s\to D_s^{\ast+}D_s^{\ast-}$, one can extract
$(\Gamma_H,
\Gamma_L, \Delta m)_{B_s}$.

A characteristic feature of the angular distributions considered in
this paper is the fact that they contain terms describing
interference effects between CP-even and CP-odd final-state  
configurations.
Because of the lifetime difference, these contributions give rise to  
a term in the time evolution of the {\it untagged} rate, which is
proportional to \cite{fd1}:
\beq\label{phickm}
\left(e^{-\Gamma_H t}-e^{-\Gamma_L
t}\right)\sin\phi_{\mbox{{\scriptsize CKM}}},
\eeq
where $\phi_{\mbox{{\scriptsize CKM}}}$ is a weak phase that is
introduced through the CKM matrix \cite{km}. In the $B_s$ decays
considered in the present paper, $\phi_{\mbox{{\scriptsize CKM}}}$
is related to the Wolfenstein parameter $\eta$ \cite{wolf}.
It is a  remarkable
feature that time-evolved {\it untagged} data
samples of angular distributions of $B_s$ decays may exhibit
CP-violating effects, if $\Delta \Gamma$ is sizeable~\cite{fd1,fd2}.
This feature may be important, because it provides an alternative to
previous investigations, which have shown how to extract
$\sin\phi_{\mbox{{\scriptsize CKM}}}$ from tagged, time-dependent  
analyses
\cite{cptaggedbs,dsnowmass93}.  This extraction, however, may not be
feasible
in the near future because it requires tagging and superb vertex
detectors, which 
must resolve the rapid $\Delta mt$ oscillations. In contrast, any
dependence on $\Delta mt$ cancels in untagged data samples, which
therefore allow feasibility studies with current vertex technology
\cite{bsbsbar}.

Concerning tests of the {\it factorization
hypothesis} \cite{fstech}--\cite{browderh}, we divide the 
$\bar b\to\bar s c\bar c$ modes into the following two
categories:
\begin{itemize}
\item colour-suppressed decays: $B_q\to J/\psi V$ with
$(q,V)\in\{(s,\phi);
(d,K^{\ast0}); (u,K^{\ast+})\}$ \cite{aleksan,bijnens}.
\item colour-allowed decays: $B_q\to D^{\ast+}_s \overline D^{\ast}_q$
with $q\in\{s,d,u\}$ \cite{rosner,mannelrr}.
\end{itemize}
Whereas the validity of the factorization assumption is very doubtful
in the colour-suppressed case, it should work much better for the
colour-allowed channels because of colour transparency~\cite{bjorken}.
The latter have furthermore rather tight
restrictions from the Heavy Quark Effective Theory (HQET) \cite{hqet}
for the form factors describing the ``factorized'' hadronic matrix
elements
of the relevant four-quark
operators~\cite{rosner,mannelrr,neubertstech}.

Our paper is organized as follows: in Section~\ref{tme} we calculate
the transition matrix elements and observables of the
angular distributions by using an appropriate low-energy effective
Hamiltonian. There we also give estimates for these observables,
allowing a comparison with experimental data.
The efficient experimental determination
of these observables is the subject of Section~\ref{ama}, where we
shall
discuss the angular moment analysis. Sections~\ref{csd} and \ref{cad}
are devoted to the angular correlations in
the colour-suppressed decays
$B_s\to J/\psi\,\phi$, $B \to J/\psi K^{\ast}$ and the colour-allowed
decays
$B_s\to D^{\ast+}_s
D^{\ast-}_s$, $B \to D^{\ast+}_s \overline D^{\ast}$, respectively.
There we
give the time evolutions of the angular distributions,
appropriate weighting functions, and discuss CP-violating effects.
Finally in Section~\ref{sum} the main results are summarized.

\section{Transition matrix elements and observables}\label{tme}
Before we present an efficient method for extracting the
observables of the angular distributions from experimental data --
the {\it angular moment analysis} -- let us discuss in this
section how these observables are calculated and what orders of
magnitude we expect for them.

\subsection{General aspects}

In order to calculate the decay amplitudes of the $\bar b\to\bar
sc\bar c$
transitions considered in this paper, we use an appropriate low-energy 
effective Hamiltonian, which has the following structure:
\beq\label{Heff}
\heff=\frac{\gf}{\sqrt{2}}\left[
\sum\limits_{j=u,c}\lambda_j^{(s)}\left\{
Q^{j}_1 C_1(\mu)+Q^{j}_2C_2(\mu)+
\sum\limits_{k=3}^{10}Q_k C_k(\mu)\right\}\right].
\eeq
Here the quantities $\lambda_j^{(s)}\equiv V_{js} V_{jb}^\ast$ denote 
CKM factors,
\begin{displaymath} 
Q_1^c = (\bar c_{\alpha} s_{\beta})_{{\rm V-A}}\;(\bar b_{\beta} 
c_{\alpha})_{{\rm V-A}}\,,
~~~~~Q_2^c = (\bar c_\alpha s_\alpha)_{{\rm V-A}}\;
(\bar b_\beta c_\beta)_{{\rm V-A}} 
\end{displaymath}
\begin{equation}\label{O1u} 
Q_1^u = (\bar u_{\alpha} s_{\beta})_{{\rm V-A}}\;(\bar b_{\beta} 
u_{\alpha})_{{\rm V-A}}\,,
~~~~~Q_2^u = (\bar u_\alpha s_\alpha)_{{\rm V-A}}\;
(\bar b_\beta u_\beta)_{{\rm V-A}} 
\end{equation}
are ``current--current'' operators,
\begin{displaymath}
Q_3 = (\bar b_\alpha s_\alpha)_{{\rm V-A}}\sum_{q=u,d,s,c,b}
(\bar q_\beta q_\beta)_{{\rm V-A}}\,,~~~~~  
Q_4 = (\bar b_{\alpha} s_{\beta})_{{\rm V-A}}\sum_{q=u,d,s,c,b}(\bar 
q_{\beta} q_{\alpha})_{{\rm V-A}} 
\end{displaymath}
\begin{equation}\label{O3}
Q_5 = (\bar b_\alpha s_\alpha)_{{\rm V-A}} 
\sum_{q=u,d,s,c,b}(\bar q_\beta q_\beta)_{{\rm V+A}}\,,~~~~~  
Q_6 = (\bar b_{\alpha} s_{\beta})_{{\rm V-A}}\sum_{q=u,d,s,c,b}
(\bar q_{\beta} q_{\alpha})_{{\rm V+A}}
\end{equation}
describe QCD penguins, while the operators
\begin{displaymath}
Q_7 = \frac{3}{2}\,(\bar b_\alpha s_\alpha)_{{\rm V-A}}\sum_{q=u,d,s,c,b}e_{q}
\,(\bar q_\beta q_\beta)_{{\rm V+A}}\,,~~~~~   
Q_8 = \frac{3}{2}\,(\bar b_{\alpha} s_{\beta})_{{\rm V-A}}
\sum_{q=u,d,s,c,b}e_{q}\,(\bar q_{\beta} q_{\alpha})_{{\rm V+A}} 
\end{displaymath}
\begin{equation}\label{EWP2}
Q_9 = \frac{3}{2}\,(\bar b_\alpha s_\alpha)_{{\rm V-A}} 
\sum_{q=u,d,s,c,b}e_{q}\,(\bar q_\beta q_\beta)_{{\rm V-A}}\,,~~~~~  
Q_{10} = \frac{3}{2}\,(\bar b_{\alpha} s_{\beta})_{{\rm V-A}}\sum_{q=u,d,s,c,b}
e_{q}\,(\bar q_{\beta} q_{\alpha})_{{\rm V-A}}
\end{equation}
are ``electroweak'' penguin operators. Here $\mbox{V}\pm\mbox{A}$ 
corresponds to $\gamma_\mu(\hat1\pm\gamma_5)$ quark currents, Greek indices 
are associated with the $SU(3)_{\mbox{{\scriptsize C}}}$ quark-colour, and 
the quantities $e_q$ arising in the expressions for the electroweak penguin
operators label the electrical quark charges. Nowadays, the Wilson 
coefficient functions $C_k(\mu)$ of the low-energy effective 
Hamiltonian $\heff$, where $\mu={\cal O}(m_b)$ denotes the usual 
renormalization scale, are known beyond the leading logarithmic 
approximation \cite{burasnlo}. 

Since $\lambda_u^{(s)}$ is suppressed with respect to $\lambda_c^{(s)}$
by a CKM factor $\lambda^2R_b$, where $\lambda=0.22$ is the
Wolfenstein
parameter \cite{wolf} and
\beq\label{rb}
R_b\equiv\frac{1}{\lambda}\frac{|V_{ub}|}{|V_{cb}|}
\eeq
is constrained by present experimental data to lie
within the range $R_b=0.36\pm0.08$ \cite{Vub, bf-rev}, and since
furthermore the current--current operators $Q^u_{1}, Q_2^u$ may  
contribute
only through penguin-like matrix elements to $\bar b\to\bar sc\bar c$
modes, the corresponding transition amplitudes are
dominated to an {\it excellent approximation} by the contribution
proportional to $\lambda_c^{(s)}$ (for a detailed discussion, 
see \cite{rf-rev}). In the penguin operators, 
we neglect the parts of flavour structure different from
$(\bar cc)(\bar bs)$.  Then the
number of relevant operators reduces from ten to four and the 
structure of the decay amplitude simplifies considerably.

In order to implement the factorization hypothesis by
factorizing the hadronic matrix elements of the four-quark
operators $Q_k$ into hadronic matrix elements of quark currents, we
have to perform suitable Fierz transformations of the operator basis
specified in (\ref{O1u})--(\ref{EWP2}). Beyond the leading logarithmic
approximation
one has to be very careful in performing such Fierz transformations,
as the Wilson coefficients depend both on the form of the chosen
operator
basis and on the applied renormalization scheme \cite{burasnlo}.
Since we do not use any specific Wilson coefficients to obtain
numerical
estimates in this paper, we may perform such Fierz transformations
and will use a tilde (\~{ }) to indicate Fierz-transformed operators. For a
discussion of the renormalization-scheme dependences arising beyond
the leading logarithmic approximation and their consistent
cancellation
in the physical transition amplitudes through certain one-loop
matrix elements at $\mu={\cal O}(m_b)$,
the reader is referred to Ref.~\cite{rf}.

Let us, in the following two subsections, investigate the structure of
the hadronic matrix elements of the low-energy effective Hamiltonian
[Eq.~(\ref{Heff})] for the exclusive colour-suppressed and  
colour-allowed
decays
$B_s\to J/\psi\,\phi$, $B \to J/\psi K^{\ast}$ and
$B_s\to D^{\ast +}_sD^{\ast-}_s$, $B \to D^{\ast +}_s \overline
D^{\ast}$,
respectively.

\subsection{Colour-suppressed decays}\label{mcsd}

If we perform a Fierz transformation of the current--current operators
specified in (\ref{O1u}), the decay amplitude for $B_q\to J/\psi
V$ ($(q,V)\in\{(s,\phi);(d,K^{\ast0});(u,K^{\ast+})\}$) 
can be written in the following form:
\begin{eqnarray}
\lefteqn{\langle J/\psi(\lambda)V(\lambda)|\heff|B_q\rangle
=\frac{\gf}{\sqrt{2}}V_{cs}V^\ast_{cb}}\label{simplema}\\
\lefteqn{~~~~~\times
\left[{\cal C}_1^{\mbox{{\scriptsize eff}}}(\mu)\langle
J/\psi(\lambda)
V(\lambda)|\tilde Q_1^c(\mu)|B_q\rangle + {\cal C}_{
\mbox{{\scriptsize 1,oct}}}^{\mbox{{\scriptsize eff}}}
(\mu)\langle J/\psi(\lambda)V(\lambda)|\tilde
Q^c_{\mbox{{\scriptsize
1,oct}}}(\mu)|B_q\rangle\right.}\nonumber\\
&&\left.+{\cal C}_5^{\mbox{{\scriptsize eff}}}(\mu)\langle
J/\psi(\lambda)
V(\lambda)|Q_5^c(\mu)|B_q\rangle + {\cal C}_{
\mbox{{\scriptsize 5,oct}}}^{\mbox{{\scriptsize eff}}}
(\mu)\langle J/\psi(\lambda)V(\lambda)|
Q^c_{\mbox{{\scriptsize
5,oct}}}(\mu)|B_q\rangle\right],\nonumber
\end{eqnarray}
where $\lambda$ denotes the helicities of the final-state vector
mesons and the ``effective'' Wilson coefficient
functions are given by
\begin{eqnarray}
{\cal C}_1^{\mbox{{\scriptsize eff}}}(\mu)~~&\equiv&C_1(\mu)+
\frac{1}{3}C_2(\mu)+C_3(\mu)+\frac{1}{3}C_4(\mu)+C_9(\mu)+
\frac{1}{3}C_{10}(\mu)\\
{\cal C}_{\mbox{{\scriptsize 1,oct}}}^{\mbox{{\scriptsize
eff}}}(\mu)&
\equiv&2\left[C_2(\mu)+C_4(\mu)+C_{10}(\mu)\right]\\
{\cal C}_5^{\mbox{{\scriptsize eff}}}(\mu)~~&\equiv&C_5(\mu)+
\frac{1}{3}C_6(\mu)+C_7(\mu)+\frac{1}{3}C_8(\mu)\\
{\cal C}_{\mbox{{\scriptsize 5,oct}}}^{\mbox{{\scriptsize
eff}}}(\mu)&
\equiv&2\left[C_6(\mu)+C_8(\mu)\right].
\end{eqnarray}
The $\mu$-dependence of these Wilson coefficients is cancelled by  
that of
the hadronic matrix elements appearing in Eq.~(\ref{simplema}).
In deriving the transition matrix element in Eq.~(\ref{simplema}),
we have used the relations
\begin{eqnarray}
\tilde Q_2^{c}&=&\frac{1}{3}\tilde Q_1^{c}+2
\tilde Q^{c}_{\mbox{{\scriptsize 1,oct}}}\\
Q_6^{c}&=&\frac{1}{3}Q_5^{c}+2
Q^{c}_{\mbox{{\scriptsize 5,oct}}}
\end{eqnarray}
with
\begin{eqnarray}
\label{q1}
\tilde Q_1^{c}&=&\left(\bar c_\alpha c_\alpha
\right)_{\mbox{{\scriptsize
V--A}}}\left(\bar b_\beta s_\beta\right)_{\mbox{{\scriptsize
V--A}}}\\
\tilde Q^{c}_{\mbox{{\scriptsize 1,oct}}}&=&\left(\bar
c_\alpha
T^a_{\alpha\beta}c_\beta\right)_{\mbox{{\scriptsize V--A}}}
\left(\bar b_\gamma T^a_{\gamma\delta}
s_\delta\right)_{\mbox{{\scriptsize
V--A}}}
\end{eqnarray}
and
\begin{eqnarray}
Q_5^{c}&=&\left(\bar c_\alpha
c_\alpha\right)_{\mbox{{\scriptsize
V+A}}}\left(\bar b_\beta s_\beta\right)_{\mbox{{\scriptsize V--A}}}\\
Q^{c}_{\mbox{{\scriptsize 5,oct}}}&=&\left(\bar c_\alpha
T^a_{\alpha\beta}c_\beta\right)_{\mbox{{\scriptsize V+A}}}
\left(\bar b_\gamma T^a_{\gamma\delta}
s_\delta\right)_{\mbox{{\scriptsize
V--A}}}.
\label{q5}
\end{eqnarray}
Here the $3\times3$ matrices $T^a$ are the $SU(3)_{\mbox{{\scriptsize C}}}$
generators, normalized to tr$(T^a\,T^b) = \delta^{ab}/2$. As we will see
below, the form of the Fierz-transformed operators given above is better
suited to analyse the $B_q\to J/\psi V$ decays since the $J/\psi$
is related to the $(\bar cc)$ pieces. The penguin contributions to
${\cal C}_1^{\mbox{{\scriptsize eff}}}(\mu)$ and 
${\cal C}_{\mbox{{\scriptsize 1,oct}}}^{\mbox{{\scriptsize eff}}}
(\mu)$ are at most ${\cal O}(10\%)$ and ${\cal O}(1\%)$,
respectively, as can be estimated from the values of
their Wilson coefficients \cite{burasnlo}.

If one assumes that $J/\psi$ emerges from the vector parts of the
$(\overline{c}c)_{V \pm A}$ quark currents appearing in the
operators in
Eqs.~(\ref{q1})--(\ref{q5}), the matrix elements of
$\tilde{Q}_{\mbox{\scriptsize 1(,oct)}}$ and
$\tilde{Q}_{\mbox{\scriptsize 5(,oct)}}$ will be equal
and the decay amplitude Eq.~(\ref{simplema})
can be simplified considerably.
Moreover, within the framework of naive {\it factorization},
we obtain (analogous for $Q_5^{c}$
and $Q^{c}_{\mbox{{\scriptsize 5,oct}}}$):
\begin{eqnarray}
\left\langle J/\psi(\lambda)V(\lambda)\left|\tilde
Q_1^{c}\right|B_q
\right\rangle_{\mbox{{\scriptsize f}}}
&=&\left\langle J/\psi(\lambda)\left|\left(\bar cc
\right)_{\mbox{{\scriptsize V--A}}}\right|0\right\rangle
\left\langle V(\lambda)\left|
\left(\bar bs\right)_{\mbox{{\scriptsize V--A}}}
\right|B_s\right\rangle\\
\left\langle J/\psi(\lambda)V(\lambda)\left|
\tilde Q^{c}_{\mbox{{\scriptsize 1,oct}}}\right|B_q\right
\rangle_{\mbox{{\scriptsize f}}}
&=&\left\langle J/\psi(\lambda)\left|\left(\bar c\,
T^a c\right)_{\mbox{{\scriptsize V--A}}}\right|0\right
\rangle\left\langle V(\lambda)\left|
\left(\bar b\, T^a s\right)_{\mbox{{\scriptsize
V--A}}}\right|B_s\right\rangle,\label{octma}
\end{eqnarray}
where summation over colour-indices is understood implicitly.
Consequently, since $J/\psi$ is a colour-singlet state, the factorized
hadronic matrix elements of the colour-octet operators given in 
Eq.~(\ref{octma}) vanish.

\subsection{Colour-allowed decays}\label{mcad}

In the case of the colour-allowed decays $B_q\to D^{\ast +}_s
\overline D^{\ast}_q$
($q\in\{u,d,s\}$), the transition amplitude can be written in a way
that
is completely analogous to Eq.~(\ref{simplema}):
\begin{eqnarray}
\lefteqn{\langle D^{\ast+}_s(\lambda) \overline
D^{\ast}_q(\lambda)|\heff|B_q\rangle
=\frac{\gf}{\sqrt{2}}V_{cs}V^\ast_{cb}}\label{ampca}\\
\lefteqn{~~~~~\times
\left[{\cal C}_2^{\mbox{{\scriptsize eff}}}(\mu)\langle D^{\ast+}_s
(\lambda)\overline D^{\ast}_q(\lambda)|Q_2^c(\mu)|B_q\rangle
+ {\cal C}_{
\mbox{{\scriptsize 2,oct}}}^{\mbox{{\scriptsize eff}}}
(\mu)\langle D^{\ast+}_s(\lambda)\overline D^{\ast}_q(\lambda)|
Q^c_{\mbox{{\scriptsize
2,oct}}}(\mu)|B_q\rangle\right.}\nonumber\\
&&\left.+{\cal C}_6^{\mbox{{\scriptsize eff}}}(\mu)\langle
D^{\ast+}_s
(\lambda)\overline D^{\ast}_q(\lambda)|\tilde
Q_6^c(\mu)|B_q\rangle
+ {\cal C}_{\mbox{{\scriptsize 6,oct}}}^{\mbox{{\scriptsize eff}}}
(\mu)\langle D^{\ast+}_s(\lambda)\overline D^{\ast}_q(\lambda)|\tilde
Q^c_{\mbox{{\scriptsize
6,oct}}}(\mu)|B_q\rangle\right].\nonumber
\end{eqnarray}
The corresponding effective Wilson coefficient functions are,
however,
very different:
\begin{eqnarray}
{\cal C}_2^{\mbox{{\scriptsize eff}}}(\mu)~~&\equiv&\frac{1}{3}
C_1(\mu)+C_2(\mu)+\frac{1}{3}C_3(\mu)+C_4(\mu)+\frac{1}{3}C_9(\mu)+
C_{10}(\mu)\\
{\cal C}_{\mbox{{\scriptsize 2,oct}}}^{\mbox{{\scriptsize
eff}}}(\mu)&
\equiv&2\left[C_1(\mu)+C_3(\mu)+C_{9}(\mu)\right]\\
{\cal C}_6^{\mbox{{\scriptsize
eff}}}(\mu)~~&\equiv&\frac{1}{3}C_5(\mu)+
C_6(\mu)+\frac{1}{3}C_7(\mu)+C_8(\mu)\\
{\cal C}_{\mbox{{\scriptsize 6,oct}}}^{\mbox{{\scriptsize
eff}}}(\mu)&
\equiv&2\left[C_5(\mu)+C_7(\mu)\right].
\end{eqnarray}
In deriving Eq.~(\ref{ampca}), we have used the relations
\begin{eqnarray}
Q_1^{c}&=&\frac{1}{3} Q_2^{c}+2
Q^{c}_{\mbox{{\scriptsize 2,oct}}}\\
\tilde Q_5^{c}&=&\frac{1}{3}\tilde Q_6^{c}+2
\tilde Q^{c}_{\mbox{{\scriptsize 6,oct}}}
\end{eqnarray}
with
\begin{eqnarray}
Q_2^{c}&=&\left(\bar c_\alpha s_\alpha
\right)_{\mbox{{\scriptsize
V--A}}}\left(\bar b_\beta c_\beta\right)_{\mbox{{\scriptsize
V--A}}}\\
Q^{c}_{\mbox{{\scriptsize 2,oct}}}&=&\left(\bar c_\alpha
T^a_{\alpha\beta}s_\beta\right)_{\mbox{{\scriptsize V--A}}}
\left(\bar b_\gamma T^a_{\gamma\delta}
c_\delta\right)_{\mbox{{\scriptsize
V--A}}}
\end{eqnarray}
and
\begin{eqnarray}
\tilde Q_6^{c}&=&-2\left(\bar c_\alpha L s_\alpha\right)
\left(\bar b_\beta R c_\beta\right)\\
\tilde Q^{c}_{\mbox{{\scriptsize 6,oct}}}&=&-2\left(\bar
c_\alpha L
T^a_{\alpha\beta}s_\beta\right)\left(\bar b_\gamma R
T^a_{\gamma\delta}
c_\delta\right).
\end{eqnarray}
Here $L$ and $R$ correspond to the Dirac structures $\hat1-\gamma_5$
and $\hat1+\gamma_5$, respectively. The $D_s^{\ast+}$
meson emerges from the $(\bar cs)$ pieces of these operators.
Since it is a {\it vector} meson, we have
\beq
\left\langle D^{\ast +}_s\left|\bar c_\alpha L s_\alpha\right|0\right
\rangle=0\,,
\eeq
and hence the {\it  factorized} matrix element of
$\tilde Q_6^{c}$ vanishes. As in Sec.~\ref{mcsd}, the
hadronic matrix elements of the colour-octet operators vanish within
the factorization approximation because of their colour-structure.

\subsection{Observables of the angular distributions}\label{observ}

The hadronic matrix element of a generic four-quark operator ${\cal  
Q}$
between
the state vectors $\langle V_1(\lambda)V_2(\lambda)|$ and
$|B_q\rangle$ has the following general Lorentz-decomposition
\cite{valencia,kp}:
\begin{eqnarray}
\lefteqn{\langle V_1(\lambda)V_2(\lambda)|{\cal
Q}|B_q\rangle=}\nonumber\\
&&\epsilon_{V_1,\mu}(\lambda)^\ast\epsilon_{V_2,\nu}(\lambda)^\ast
\left[a g^{\mu\nu} + \frac{b}{m_{V_1} m_{V_2}}p_{V_2}^\mu
p_{V_1}^\nu+
i\frac{c}{m_{V_1} m_{V_2}}\varepsilon^{\mu\nu\alpha\beta}
p_{V_1,\alpha}p_{V_2,\beta}\right],
\end{eqnarray}
where the symbols $\epsilon(\lambda)$ denote the polarization vectors of
the final-state vector mesons $V_1$ and $V_2$. A similar parametrization 
can be employed to express the transition matrix elements
[Eqs.~(\ref{simplema})
and (\ref{ampca})], yielding
\begin{eqnarray}
\lefteqn{a=\frac{\gf}{\sqrt{2}}V_{cs}V_{cb}^\ast\left[{\cal C}_i^
{\mbox{{\scriptsize eff}}}(\mu)A_i^{\mbox{{\scriptsize f}}}
+{\cal C}_{i+4}^{\mbox{{\scriptsize eff}}}(\mu)
A_{i+4}^{\mbox{{\scriptsize f}}}\right.}\label{aaa}\\
&&\left.+\,{\cal C}_i^{\mbox{{\scriptsize
eff}}}(\mu)A_i^{\mbox{{\scriptsize nf}}}(\mu)+
{\cal C}_{\mbox{{\scriptsize $i$,oct}}}^{\mbox{{\scriptsize
eff}}}(\mu)
A^{\mbox{{\scriptsize nf}}}_{\mbox{{\scriptsize $i$,oct}}}(\mu)
+{\cal C}_{i+4}^{\mbox{{\scriptsize
eff}}}(\mu)A_{i+4}^{\mbox{{\scriptsize
nf}}}(\mu)+
{\cal C}_{\mbox{{\scriptsize $i+4,$oct}}}^{\mbox{{\scriptsize
eff}}}(\mu)
A^{\mbox{{\scriptsize nf}}}_{\mbox{{\scriptsize $i+4$,oct}}}(\mu)
\right]\nonumber
\end{eqnarray}
\begin{eqnarray}
\lefteqn{b=\frac{\gf}{\sqrt{2}}V_{cs}V_{cb}^\ast\left[{\cal C}_i^
{\mbox{{\scriptsize eff}}}(\mu)B_i^{\mbox{{\scriptsize f}}}
+{\cal C}_{i+4}^{\mbox{{\scriptsize eff}}}(\mu)
B_{i+4}^{\mbox{{\scriptsize f}}}\right.}\label{bbb}\\
&&\left.+\,{\cal C}_i^{\mbox{{\scriptsize
eff}}}(\mu)B_i^{\mbox{{\scriptsize nf}}}(\mu)+
{\cal C}_{\mbox{{\scriptsize $i$,oct}}}^{\mbox{{\scriptsize
eff}}}(\mu)
B^{\mbox{{\scriptsize nf}}}_{\mbox{{\scriptsize $i$,oct}}}(\mu)
+{\cal C}_{i+4}^{\mbox{{\scriptsize
eff}}}(\mu)B_{i+4}^{\mbox{{\scriptsize
nf}}}(\mu)+
{\cal C}_{\mbox{{\scriptsize $i+4$,oct}}}^{\mbox{{\scriptsize
eff}}}(\mu)
B^{\mbox{{\scriptsize nf}}}_{\mbox{{\scriptsize $i+4$,oct}}}(\mu)
\right]\nonumber
\end{eqnarray}
\begin{eqnarray}
\lefteqn{c=\frac{\gf}{\sqrt{2}}V_{cs}V_{cb}^\ast\left[{\cal C}_i^
{\mbox{{\scriptsize eff}}}(\mu)C_i^{\mbox{{\scriptsize f}}}
+{\cal C}_{i+4}^{\mbox{{\scriptsize eff}}}(\mu)
C_{i+4}^{\mbox{{\scriptsize f}}}\right.}\label{ccc}\\
&&\left.+\,{\cal C}_i^{\mbox{{\scriptsize
eff}}}(\mu)C_i^{\mbox{{\scriptsize nf}}}(\mu)+
{\cal C}_{\mbox{{\scriptsize $i$,oct}}}^{\mbox{{\scriptsize
eff}}}(\mu)
C^{\mbox{{\scriptsize nf}}}_{\mbox{{\scriptsize $i$,oct}}}(\mu)
+{\cal C}_{i+4}^{\mbox{{\scriptsize
eff}}}(\mu)C_{i+4}^{\mbox{{\scriptsize
nf}}}(\mu)+
{\cal C}_{\mbox{{\scriptsize $i+4$,oct}}}^{\mbox{{\scriptsize
eff}}}(\mu)
C^{\mbox{{\scriptsize nf}}}_{\mbox{{\scriptsize $i+4$,oct}}}(\mu)
\right],\nonumber
\end{eqnarray}
where the index $i$ distinguishes between colour-suppressed ($i=1$)
and
colour-allowed ($i=2$) decays and
``f'' and ``nf'' correspond to ``factorized'' and ``non-factorized''
matrix elements, respectively. Note that the factorized amplitudes
do not depend on the renormalization scale $\mu$. Since the Wilson
coefficients depend on this scale, this already signals the need for
non-factorizable contributions to cancel the $\mu$-dependence in
Eqs.~(\ref{aaa})--(\ref{ccc}) (see e.g.\ Ref.~\cite{Buras-nf} for a
further discussion of that point).

In the following sections we will analyse the decays $B_q\to V_1V_2$
in terms of {\it  linear polarization states}. The corresponding  
decay amplitudes take the form \cite{rosner,ddlr}
\beq\label{ampl}
A(B_q(t) \to V_1V_2) =
\frac{A_0(t)}{x}  {\bep}^{*L}_{V_1}
{\bep}^{*L}_{V_2} -
A_{\|}(t) {\bep}^{*T}_{V_1} \cdot
{\bep}^{*T}_{V_2} / \sqrt{2}  -
i A_{\perp}(t) {\bep}^*_{V_1} \times
{\bep}^*_{V_2} \cdot \hat{\bf p}_{V_2} /
                    \sqrt{2}~,
\eeq
where $x\equiv p_{V_1}\cdot p_{V_2}/(m_{V_1} m_{V_2})$ and
$\hat{\bf p}_{V_2}$ is the unit vector along the direction of
motion of $V_2$ in the rest frame of $V_1$.
Here the time dependences originate from $B_q$--$\overline{B}_q$
mixing. In our notation,
an unmixed $B_q$ meson is present at $t=0$.

The linear polarization amplitudes at $t=0$ defined by  
Eq.~(\ref{ampl})
can be expressed in terms of $a$, $b$ and $c$ as follows \cite{ddlr}:
\begin{eqnarray}
A_0(0)&=&-x a - (x^2-1)b\nonumber\\
A_\parallel(0)&=&\sqrt{2}\,a\label{obs}\\
A_\perp(0)&=&\sqrt{2(x^2-1)}\,\;c.\nonumber
\end{eqnarray}
At time $t=0$, the angular distributions for $B_q\to V_1V_2$ depend on
the observables $|A_0(0)|$, $|A_\parallel(0)|$, $|A_\perp(0)|$ and
on the two phases $\delta_1\equiv\mbox{Arg}\left[A_\parallel(0)^\ast
A_\perp(0)\right]$ and $\delta_2\equiv\mbox{Arg}\left[A_0(0)^\ast
A_\perp(0)\right]$, which are CP-conserving strong phases that are
$0~(\mbox{mod}~\pi)$ in the absence of final-state interactions
(probably
not a justifiable assumption for the colour-suppressed modes).  
Quantitative
estimates
for these observables will be given in the following subsection.

\subsection{Factorization tests and estimates of
observables}\label{estimates}

While the non-factorizable contributions to $a$, $b$ and $c$ cannot
be calculated at present, the evaluation of the factorizable
contributions is straightforward.
Without yet going into the details of which form factors to employ,
the naive factorization assumption yields many testable consequences.
For example, time-reversal invariance forces the form factors
parametrizing quark currents to be all relatively real. Consequently,
naive factorization predicts the same strong phase (mod $\pi$) for  
the
three amplitudes $A_0 (0), A_{\parallel}(0), A_\perp (0)$. It  
therefore
predicts vanishing values of the two observables
\cite{kornerg,yamamoto,browderh}
\begin{equation}
\mbox{ Im }[A^*_0 (0) A_\perp (0) ]=0 
\label{factest1}
\end{equation}
\begin{equation}
\mbox{ Im }[A^*_{\parallel}(0) A_\perp (0)]=0 \,,
\label{factest2}
\end{equation}
and the equality
\begin{equation}
\mbox{ Re }[A^*_0 (0) A_{\parallel} (0)] = \pm |A_0 (0)\;
A_{\parallel} (0)|\,.
\label{factest0par}
\end{equation}
The breakdown of the naive factorization assumption is unequivocally
proved if any of the three equations  
(\ref{factest1})--(\ref{factest0par})
is not satisfied.
Detailed comparisons of polarization amplitudes in non-leptonic and  
semi-leptonic decays   test additional implications of the naive  
factorization assumption. The phenomenology of detailed studies of  
the full non-trivial angular distributions is thus much richer than  
the single factorization test available for a pseudoscalar decaying 
into two pseudoscalars \cite{bjorken}. While the above 
equations represent general tests of
the factorization assumption, it is also useful to examine the  
predictions for
the observables of the angular distributions for various form factor
ans\"{a}tze.

\subsubsection{The colour-suppressed decays $B_q\to J/\psi V$}

The factorized amplitudes for $B_q\to J/\psi V$ with
$(q,V)\in\{(s,\phi);(d,K^{\ast0});(u,K^{\ast+})\}$
are given by~\cite{bsw,aleksan,bijnens}:
\begin{eqnarray}
A_1^{\mbox{{\scriptsize f}}}&=&-\,f_{J/\psi} m_{J/\psi} (m_{B_q}+m_V)
A_1^{B_qV}(m_{J/\psi}^2)~=~A_5^{\mbox{{\scriptsize f}}}\nonumber\\
B_1^{\mbox{{\scriptsize f}}}&&=~~~~2\,\frac{f_{J/\psi} m_{J/\psi}^2 m_V}
{m_{B_q}+m_V}A_2^{B_qV}
(m_{J/\psi}^2)~~~=~~~B_5^{\mbox{{\scriptsize f}}}
\label{abcf}\\
C_1^{\mbox{{\scriptsize f}}}&&=~~~~2\,\frac{f_{J/\psi} m_{J/\psi}^2 m_V}
{m_{B_q}+m_V}V^{B_qV}
(m_{J/\psi}^2)~~~=~~~C_5^{\mbox{{\scriptsize f}}},\nonumber
\end{eqnarray}
where we have used the notation of Bauer, Stech and Wirbel for the
form factors $A_i^{B_qV}(q^2)$ and $V^{B_qV}(q^2)$ of quark currents
\cite{bsw}. The parameter $f_{J/\psi}$ denotes the $J/\psi$ decay
constant, which can be determined from the $J/\psi\to e^+e^-$ rate, 
yielding $f_{J/\psi}=395\,\mbox{MeV}$.

At present, several methods for obtaining the form factors
$A_1(m_{J/\psi}^2)$,
$A_2(m_{J/\psi}^2)$ and $V(m_{J/\psi}^2)$ for the $B\to K^\ast$ case
are on the market. Using $SU(3)$ flavour symmetry of strong
interactions,
the $B\to K^\ast$ form factors can be related to the $B_s\to\phi$ case. 
In Table~\ref{predictions} we have collected the form factors proposed
by several authors \cite{bsw,soares,cheng}, and have
moreover given the corresponding predictions for the {\it  ratios} of
observables of the angular distributions. These ratios should suffer
less from unknown $SU(3)$-breaking corrections than the  
observables themselves. Note that these ratios are independent of the 
Wilson coefficients within the factorization approach.

The quantity
\beq
\frac{\Gamma_0(0)}{\Gamma_0(0)+\Gamma_T(0)}\equiv\frac{|A_0(0)|^2}
{|A_0(0)|^2+|A_\parallel(0)|^2+|A_\perp(0)|^2}
\eeq
describes the ratio of the longitudinal to the total rate at $t=0$.
Although CDF \cite{cdf} claims to have measured this quantity, from
their untagged data sample, to be $0.56\pm0.21
(\mbox{stat.})^{+0.02}_{-0.04}(\mbox{syst.})$, their
claim is valid only if the CP-odd component of $B_s\rightarrow
J/\psi\phi$ is negligible, or if the lifetime difference 
$\Delta\Gamma$ can be ignored.

\begin{table}[t]
\begin{center}
\begin{tabular}{|c|c|c|c|}
\hline
Observable & BSW \cite{bsw} & Soares \cite{soares} & Cheng
\cite{cheng} \\
\hline
$A_1^{BK^\ast}(m_{J/\psi}^2)$ & 0.46 & 0.42 & 0.41 \\
\hline
$A_2^{BK^\ast}(m_{J/\psi}^2)$ & 0.46 & 0.43 & 0.36 \\
\hline
$V^{BK^\ast}(m_{J/\psi}^2)$   & 0.55 & 1.08 & 0.72 \\
\hline
$|A_\parallel(0)|/|A_0(0)|$ & 0.81 (0.77) & 0.82 (0.78)& 0.75  
(0.70)\\
\hline
$|A_\perp(0)|/|A_0(0)|$ & 0.41 (0.40)& 0.89 (0.88)& 0.55 (0.54)\\
\hline
$\Gamma_0(0)/(\Gamma_0(0)+\Gamma_T(0))$ & 0.55 (0.57)& 0.40 (0.42)&
0.54 (0.56) \\
\hline
$\delta_1$ & $\pi$ & $\pi$ & $\pi$ \\
\hline
$\delta_2$ &  0    &    0  &   0  \\
\hline
\end{tabular}
\end{center}
\caption{Predictions for form factors and $B_s\to J/\psi \phi$ ($B \to
J/\psi K^{\ast}$) observables.}\label{predictions}
\end{table}

The $2^{\mbox{{\scriptsize nd}}}$--$4^{\mbox{{\scriptsize th}}}$
columns of Table~\ref{predictions} are calculated within the  
framework of naive factorization, i.e.\ we have inserted 
Eq.~(\ref{abcf}) into Eqs.~(\ref{aaa})--(\ref{ccc}) and have 
omitted the ``nf'' terms in order to calculate 
the amplitudes in Eq.~(\ref{obs}).

The form factors given by Soares \cite{soares} are obtained from
$\overline D\to K^{(\ast)}l\overline{\nu}_l$ data by
using heavy-quark symmety relations \cite{iw} and assuming the
monopole momentum-transfer dependence of the BSW model \cite{bsw}.
Some more form-factor models and their predictions are discussed
in \cite{cheng2}. Note that the small difference between the
$B_s\to J/\psi\,\phi$ and $B \to J/\psi K^{\ast}$ results in
Table~\ref{predictions} is related to phase-space effects and not  
to any $SU(3)$-breaking effects in the corresponding hadronic 
matrix elements.

Looking at Table~\ref{predictions}, we observe that the  
``factorized'' predictions for $|A_\parallel(0)|/|A_0(0)|$ are 
rather stable $(\approx0.8)$, while $|A_\perp(0)|/|A_0(0)|$ 
depends strongly on the method used for obtaining the 
form factors. A common feature of all results is $\delta_1=\pi$ 
and $\delta_2=0$. Therefore a measurement of non-trivial phases 
$\delta_1$ and $\delta_2$ would imply the presence of strong 
final-state interactions and non-factorizable contributions.

Whereas the use of the factorization assumption is very questionable
in the case of the channels $B_s\to J/\psi\,\phi$ and $B \to 
J/\psi K^\ast$, flavour $SU(3)$ symmetry is probably a good working 
assumption. Thus all the hadronization dynamics of the 
$B_s\rightarrow J/\psi\phi$ decay, such as the phases $\delta_1$ 
and $\delta_2$ and magnitudes of the amplitudes
\begin{equation}
A_0 (0), \;A_{\parallel}(0),\;A_\perp (0),
\end{equation}
can be obtained from the $B\rightarrow J/\psi K^*$  
modes.\footnote{Although those $SU(3)$ relations are mostly 
trivial, one subtlety due to quantum-coherence must be
emphasized. Because of the $SU(3)$ relations in the unmixed 
amplitudes
$$
A_f(B_s \to J/\psi \phi) = A_f(B \to J/\psi K^*)~, ~~~\mbox{where}~~
f=0, \|, \perp ~~,
$$
the magnitudes of the amplitudes for $B_s^L$ or $B_s^H$ decays
into CP-even or CP-odd $J/\psi\phi$ final-state configurations,
respectively, are a factor of $\sqrt{2}$ larger than their 
corresponding $B\rightarrow J/\psi K^*$ ones. [Here the $K^*$ is 
seen in a flavour-specific mode.  If $K^*$ is neutral and is 
observed as $\pi^0 K_S$, quantum coherence in $B^0-\overline{B^0}$ 
must also be taken into account.]  If the CP-even processes 
dominate, then
$$
\Gamma(B_s^L \rightarrow J/\psi\phi ) \approx
2\Gamma(B\rightarrow J/\psi K^* )~~.
$$
Studies of $B_s$ versus $B$ production fractions can thus be  
undertaken, since the lifetimes will be precisely known.}
This approach may be helpful to extract the CKM phase 
$\phi_{\mbox{{\scriptsize CKM}}}$ (see Eq.~(\ref{phickm})), 
as we will see below.

The factorization assumption should work much better for the
transitions $B_s\to D_s^{\ast+}D_s^{\ast-}$ and $B \to
D_s^{\ast+} \overline D^{\ast}$.
Therefore the results presented in the following subsection 
should be more reliable than those summarized in 
Table~\ref{predictions}.

\subsubsection{The colour-allowed decays $B_q\to D^{\ast+}_s \overline
D^{\ast}_q$}

Using again the same notation as Ref.~\cite{bsw}, we get the
following
``factorized'' results for the modes $B_q\to D^{\ast +}_s \overline
D^{\ast}_q$
($q\in\{u,d,s\}$)~\cite{rosner,mannelrr,neubertstech}:
\begin{eqnarray}
A_2^{\mbox{{\scriptsize f}}}&=&-f_{D^{\ast}_s} m_{D^{\ast}_s}
(m_{B_q}+m_{D^{\ast}_q})A_1^{B_qD^{\ast}_q}(m_{D^{\ast}_s}^2),~~
A_6^{\mbox{{\scriptsize f}}}=0\nonumber\\
B_2^{\mbox{{\scriptsize f}}}&=&2\frac{f_{D^{\ast}_s}
m_{D^{\ast}_s}^2 m_{D^{\ast}_q}}
{m_{B_q}+m_{D^{\ast}_q}}A_2^{B_qD^{\ast}_q}(m_{D^{\ast}_s}^2),
\quad~~~~~~~~~~~
B_6^{\mbox{{\scriptsize f}}}=0\label{abcfca}\\
C_2^{\mbox{{\scriptsize f}}}&=&2\frac{f_{D^{\ast}_s} m_{D^{\ast}_s}^2
m_{D^{\ast}_q}}
{m_{B_q}+m_{D^{\ast}_q}}V^{B_qD^{\ast}_q}(m_{D^{\ast}_s}^2),
\quad~~~~~~~~~~~
C_6^{\mbox{{\scriptsize f}}}=0.\nonumber
\end{eqnarray}
The parameter $f_{D^\ast_s}$ is the $D^\ast_s$ decay constant. The
spin
symmetry of HQET implies $f_{D_s^\ast}\approx f_{D_s}$. A recent  
compilation
of measurements of $f_{D_s}$ from $D_s \to \mu \overline\nu$ gives 
$(241 \pm 21 \pm 30)$ MeV~\cite{fdsrichman}.

In the case of $B_q\to \overline D_q^\ast$ transitions we have 
rather tight restrictions from HQET (for reviews, see for example
\cite{hqet}) for the corresponding form factors. The following 
ratios turn out to be useful to implement these HQET 
constraints
\cite{neubert-rev}:
\begin{eqnarray}
R_1(w)&=&\left[1-\frac{q^2}{\left(m_{B_q}+m_{D_q^\ast}\right)^2}
\right]
\frac{V^{BD^\ast}(q^2)}{A_1^{BD^\ast}(q^2)}\\
R_2(w)&=&\left[1-\frac{q^2}{\left(m_{B_q}+m_{D_q^\ast}\right)^2}
\right ]
\frac{A_2^{BD^\ast}(q^2)}{A_1^{BD^\ast}(q^2)},
\end{eqnarray}
where $R_1(w)$ and $R_2(w)$ are defined in such a way that we have
\beq
R_1(w)=R_2(w)=1
\eeq
for all values of $w$ in the {\it strict} heavy-quark limit. The
kinematical variable $w$ is defined by
\beq
w=\frac{m_{B_q}^2+m_{D_q^\ast}^2-q^2}{2m_{B_q}m_{D_q^\ast}}.
\eeq
The value of the momentum transfer $q^2$ relevant for
Eq.~(\ref{abcfca}) is $q^2=m_{D_s^\ast}^2$. The form factor 
$A_1^{BD^\ast}(q^2)$ is usually written as
\beq
A_1^{BD^\ast}(q^2)=\frac{m_{B_q}+m_{D_q^\ast}}{2\sqrt{m_{B_q}
m_{D_q^\ast}}}\left[1-\frac{q^2}{\left(m_{B_q}+m_{D_q^\ast}\right)^2}
\right]
h_{A_1}(w),
\eeq
where $h_{A_1}(w)$ corresponds to the Isgur--Wise function
in the strict heavy-quark limit and can be written as
\beq
h_{A_1}(w)={\cal F}(1)\left[1-\rho^2_{A_1}(w-1)+{\cal
O}((w-1)^2)\right].
\eeq
The current status of the normalization ${\cal F}(1)$ and of the
``slope parameter'' $\rho^2_{A_1}$ has been summarized recently by Neubert
in Ref.~\cite{neubert-rev}. The form factor $A_1^{BD^\ast}(q^2)$ is
protected by Luke's theorem \cite{luke} against $1/m_Q$ 
corrections at zero recoil. The other form factors 
$A_2^{BD^\ast}(q^2)$ and $V^{BD^\ast}(q^2)$ are not protected by 
this theorem. From calculations based on HQET one expects a rather 
weak dependence of $R_1(w)$ and $R_2(w)$ on $w$ and therefore uses
\begin{eqnarray}
R_1(w)&=&R_1[1+{\cal O}(w-1)]\\
R_2(w)&=&R_2[1+{\cal O}(w-1)].
\end{eqnarray}
In our analysis we will neglect the $w$-dependence completely.

\begin{table}[t]
\begin{center}
\begin{tabular}{|c|c|c|c|}
\hline
Observable & BSW  & $\mbox{HQET}_{\mbox{{\scriptsize strict}}}$ &
$\mbox{HQET}_{\mbox{{\scriptsize sym..-break.}}}$\\
\hline
$A_1^{BD^\ast}(m_{D_s^{\ast}}^2)$ & 0.72 & 0.70 (0.68) & 0.70
(0.68)\\
\hline
$A_2^{BD^\ast}(m_{D_s^{\ast}}^2)$ & 0.76 & 0.76 (0.75) & 0.54
(0.53)\\
\hline
$V^{BD^\ast}(m_{D_s^{\ast}}^2)$   & 0.79 & 0.76 (0.75) & 0.89
(0.88)\\
\hline
$|A_\parallel(0)|/|A_0(0)|$ & 0.90 (0.90) & 0.91 (0.91) & 0.81 (0.80)
\\
\hline
$|A_\perp(0)|/|A_0(0)|$ & 0.32 (0.32)& 0.32 (0.33) & 0.33 (0.34)\\
\hline
$\Gamma_0(0)/(\Gamma_0(0)+\Gamma_T(0))$ & 0.52 (0.52)& 0.52 (0.52) &
0.57 (0.57)\\
\hline
$\delta_1$ & $\pi$ & $\pi$ &$\pi$\\
\hline
$\delta_2$ &  0    &    0   &   0\\
\hline
\end{tabular}
\end{center}
\caption{Predictions for form factors and $B_s\to
D_s^{\ast+}D_s^{\ast-}$
($B \to D_s^{\ast+}\overline D^{\ast}$)
observables.}\label{predictionsca}
\end{table}

Following these lines we have calculated the results for the
form factors and ratios of observables, which should receive 
smaller $SU(3)$-breaking corrections than the observables 
themselves, summarized in Table~\ref{predictionsca}. For 
completeness we have also given the results obtained by applying 
the BSW model \cite{bsw} in the $2^{\mbox{{\scriptsize nd}}}$ column. 
In order to calculate the $3^{\mbox{{\scriptsize rd}}}$ and 
$4^{\mbox{{\scriptsize th}}}$ columns, we have used 
${\cal F}(1)=0.91$ and $\rho^2_{A_1}=0.91$ \cite{neubert-rev}. 
The columns denoted by
$\mbox{HQET}_{\mbox{{\scriptsize strict}}}$ and
$\mbox{HQET}_{\mbox{{\scriptsize sym.-break.}}}$ correspond to
$R_1=R_2=1$ and $R_1=1.18$, $R_2=0.71$, respectively, where we have
employed the results by Neubert \cite{neubert-rev}
to take into account HQET symmetry-breaking corrections. Within 
the factorization approximation we obtain the following simple 
expressions for $A_\parallel(0)/A_0(0)$ and $A_\perp(0)/A_0(0)$
in terms of the HQET parameters:
\begin{eqnarray}
\frac{A_\parallel(0)}{A_0(0)}&=&\sqrt{2}\left[
\frac{m_{D_s^\ast}}{m_{B_q}}\left(\frac{x^2-1}{w+1}\right)
R_2(w)-x\right]^{-1}\\
\frac{A_\perp(0)}{A_0(0)}&=&
-\left[\frac{m_{D_s^\ast}}{m_{B_q}}\left(\frac{\sqrt{x^2-1}}{w+1}
\right)
R_1(w)\right]\frac{A_\parallel(0)}{A_0(0)},
\end{eqnarray}
where the kinematical variable
\beq
x=\frac{m_{B_q}^2-m_{D_q^\ast}^2-m_{D_s^\ast}^2}{2m_{D_q^\ast}
m_{D_s^\ast}}
\eeq
has been defined after Eq.~(\ref{ampl}).

If we compare Table~\ref{predictionsca} with 
Table~\ref{predictions}, we note that the results for the 
observables depend much less on the way of obtaining the 
form factors. Also the ``old'' BSW model is in rather 
good agreement with the HQET predictions, which is quite 
remarkable. Therefore the results given in
Table~\ref{predictionsca} are more reliable than those 
collected in Table~\ref{predictions}. In this respect it is 
also important to note that the non-factorizable
contributions appearing in Eqs.~(\ref{aaa})--(\ref{ccc}) should 
play a minor role for the colour-allowed decay class and that 
$\delta_1=\pi$ and $\delta_2=0$ is expected to hold on rather 
solid ground. Because of the latter feature, CP-violating effects 
arising in untagged $B_s\to D_s^{\ast+}D_s^{\ast-}$ data samples 
should be a promising way to extract the weak phase 
$\phi_{\mbox{{\scriptsize CKM}}}$ (see Eq.~(\ref{phickm})),
as has been outlined in detail in Ref.~\cite{fd1}.

\section{The angular-moment analysis}\label{ama}

The main focus of this section is the efficient determination of 
the observables discussed in Sect.\ \ref{observ} and 
\ref{estimates}. This can be accomplished by an {\it  
angular-moment analysis} \cite{dqstl}. In this approach, the
observed data are weighted by judiciously chosen functions, which
project out any desired observable. Whereas Ref.~\cite{dqstl} 
determines the moments for a few choice angular
distributions, using spherical harmonics, this paper indicates how 
to determine suitable weighting functions for all kinds of angular 
distributions, using only orthogonality arguments (without invoking 
spherical harmonics).

\bigskip

Let us denote the angular distribution of a given decay by
\beq
f(\Theta, \alpha; t) =
        \sum_{i} b^{(i)}(\alpha; t) g^{(i)}(\Theta),
\label{f=bg}
\eeq
where ${\bf \alpha}$ represents all the parameters that are  
independent of the kinematics, which is described by certain 
decay angles. In general, the physical process involves an 
arbitrary number of such angles denoted generically by $\Theta$.
For the examples considered in this article, $i$ runs from 1 to 
6, and we have
$${\bf \alpha} =
\{ \Gamma_H, \Gamma_L, \Delta m, |A_0(0)|, |A_{\|}(0)|,
|A_{\perp}(0)|,
\delta_1, \delta_2,\phi_{\mbox{{\scriptsize CKM}}} \}.$$
All the quantities of interest are encoded in the time evolution 
of the observables $b^{(i)}(\alpha;t)$. In the following discussion 
the $\alpha$- and $t$-dependence of the $b^{(i)}$'s is implicit 
wherever not explicitly stated.

\medskip

The usual method for extracting $b^{(i)}$'s is to
use an  {\it  unbinned maximum likelihood fit} \cite{fit}.
Performing such a fit for a given quantity requires some
idea of the values of the other quantities.
When one deals with limited statistics, one may want to exploit
alternative methods that completely decouple the extraction of 
one observable from all the others. Luckily such a
method exists, the angular-moment analysis.

\medskip

If we can find a weighting function $w^{(i)}(\Theta)$
for each $i$ such that
\beq
\int [D\Theta] w^{(i)}(\Theta)
        g^{(k)}(\Theta) =
        \delta_{ik} =
                \left\{ \begin{array}{ll}
                0 & \mbox{for $i \neq k$} \\
                1 & \mbox{for $i = k$,}
                \end{array}
        \right. ~~~
\eeq                    
then the $b^{(i)}$'s can be obtained directly from
\beq
b^{(i)} = \int [D\Theta]
        w^{(i)}(\Theta) f_{{\mbox{{\scriptsize
expt}}}}
(\Theta)                ~.
\eeq
Here $[D\Theta]$ denotes the appropriate measure for integrating over 
all angles $\Theta$, and
$f_{{\mbox{{\scriptsize expt}}}}(\Theta)$
denotes the observed full angular distribution.
For a small number of events $(N)$, the form of the function
$f_{{\mbox{{\scriptsize expt}}}}(\Theta)$ will not be known, 
but only the values for $\Theta$ will be known for each event.
In that case the above equation reduces to
\beq
b^{(i)}= \frac{1}{N}
\sum_{\mbox{{\scriptsize events}}} w^{(i)}(\Theta)~.
\eeq
These $b^{(i)}$'s can then be used directly for studying their
$(\alpha; t)$ dependence.

That $w^{(k)}$'s can always be found for any angular distribution
follows from the linear independence of the $g^{(i)}$'s (they 
have to be independent for the angular distribution to be 
legitimate). The vector space ${\cal V}_k$ spanned by all 
$g^{(j)}$'s for $j \neq k$ is a {\it  proper} subspace of the 
vector space ${\cal V}$ spanned by all $g^{(i)}$'s.
Then there exists a one-dimensional vector space ${\cal W}$ such that
${\cal V} = {\cal V}_k \oplus {\cal W}$ and ${\cal V}_k \perp {\cal
W}$. Here the scalar product is defined as
$v_1 \cdot v_2 \equiv \int [D\Theta]  v_1(\Theta)v_2(\Theta)$;
$w^{(k)}$ is then the element of ${\cal W}$ with proper magnitude.

For a given set of $g^{(i)}$'s, the choice of
$w^{(i)}$'s need not be unique. We can
always take any vector space ${\cal V}' \supset {\cal V}$ and
the corresponding {\it  projection} space ${\cal W}'$ such that
${\cal V}' = {\cal V}_k \oplus {\cal W}'$ and
${\cal V}_k \perp {\cal W}'$. Then any $w^{(k)} \in {\cal W}'$ 
with $w^{(k)} \cdot g^{(k)} = 1$ will serve our purpose.

We now indicate an explicit procedure for finding a set of 
weighting functions applicable to any given angular distribution.
For a theoretical angular distribution of the form 
$f = \sum\limits_{i=1}^n b^{(i)}g^{(i)}$ (the dependence on angles and time
is implicit),
\beq
w^{(i)} = \sum_{j = 1}^n \lambda_{ij} g^{(j)}
\eeq
is a proper weighting function, where the $n^2$ unknowns $\lambda_{ij}$
are solutions of the $n^2$ simultaneous equations
\beq
\delta_{ik} = \sum_{j=1}^n \lambda_{ij} \int [D\Theta] g^{(j)}  
g^{(k)}.
\eeq
The existence of such a solution follows from the vector-space
arguments given earlier. The $w^{(i)}$'s need not be restricted 
to the vector space spanned by the $n$ vectors $g^{(j)}$'s, in 
which case the unknowns $\lambda_{ij}$ will be underdetermined and 
more than one set of $w^{(i)}$'s will serve our purpose.

It is crucial to observe that the weighting functions $w^{(i)}$
depend only on the angular terms and not on the values of the
observables $b^{(j)}$. The implication is that no matter how 
complicated the detailed angular distribution, there always 
exists an angular weighting, which projects out the desired 
observables alone. We therefore recommend the use of moments
whenever one wishes to extract observables from measured angular
distributions, such as in weak decays of baryons~\cite{kraemer} or
pseudoscalars [$P \to V \ell \nu, X_J  \ell \nu, V V,$ etc.], 
or strong and electromagnetic decays \cite{chinese}.  
The utility of this approach cannot be overemphasized.  
For instance, the moment analysis allows
the study of the $q^2$ dependence of each of the observables
separately in the process $P \to V \ell \nu$.  This could prove
useful for the extraction of form factors and the determination of
CKM elements, e.g.\ $V_{cb}$ and $V_{ub}$.

We note that there exist many legitimate choices of weighting  
functions.  The {\it optimal} choice depends on the numerical values 
of the observables~\cite{dqstl} and on the detector configuration.

\section{The angular distribution of the colour-suppressed decays
$B_s\to J/\psi\,\phi$ and $B \to J/\psi K^{\ast}$}\label{csd}
\label{Bspsiphi}

In this section we give the angular distribution of the decays
$B_s\to J/\psi\,\phi$ and $B \to J/\psi K^{\ast}$, their
time-dependences and appropriate weighting functions.

\subsection{The decay $B_s\to J/\psi(\to l^+l^-)\phi(\to K^+K^-)$}

An analysis of this process has been performed in \cite{ddlr} in terms of
linear polarization states of the final-state vector mesons. The
corresponding decay amplitude has the same form as Eq.~(\ref{ampl}).
Since the amplitudes $A_{0,\|}$ and $A_{\perp}$ are related to
CP-even and CP-odd final-state configurations, respectively, they differ
in time evolution as well as angular distribution.
The angular distribution can be used to separate these components
and their time evolution can be studied individually.

The differential decay rate at time $t$ as a function of a generic
variable $x$ will be denoted by
\beq
\frac{d\Gamma(t,x)}{d\,x}\equiv\frac{1}{N(t)}\frac{d^2 N(t)}{dxdt}.
\eeq
Consequently the normalized number of decays in the intervals
$[t,t+\Delta t]$ and $[x,x+\Delta x]$ is given by
\beq
\frac{d\Gamma(t,x)}{dx}\Delta x\Delta t=\frac{1}{N(t)}\frac{d^2
N(t)}{dxdt}
\Delta x\Delta t.
\eeq

\subsection{Tagged decays }\label{tgd}

In the case of $B_s\to J/\psi\,\phi$, the three-angle distribution for
the decay of an initially present (i.e.\ tagged) $B_s$ meson takes
the form \cite{ddlr}
$$
\frac{d^3 \Gamma [B_s(t) \to J/\psi (\to l^+ l^-) \phi (\to K^+
K^-)]}
{d \cos \theta~d \varphi~d \cos \psi}
\propto \frac{9}{32 \pi} \Bigl[~2 |A_0(t)|^2 \cos^2 \psi (1 - \sin^2
\theta
\cos^2 \varphi)
\hfill{ }
$$
$$
+ \sin^2 \psi \{ |A_\parallel(t)|^2  (1 - \sin^2 \theta \sin^2
\varphi)
+ |A_\perp(t)|^2 \sin^2 \theta - \mbox{ Im }(A_\parallel^*(t)
A_\perp(t))
\sin 2 \theta \sin \varphi \}~~~
$$
\beq \label{triple}
+\frac{1}{\sqrt{2}}\sin2\psi\{{\mbox{ Re }}(A_0^*(t) A_\parallel(t))
\sin^2
\theta
\sin2\varphi+\mbox{ Im }(A_0^*(t)A_\perp(t))\sin2\theta\cos\varphi~\}
~\Bigr]~~.
\eeq
Throughout this section we will apply the same conventions as 
in Ref.~\cite{ddlr}, i.e.\ $\phi$ moves in $x$ direction in 
the $J/\psi$ rest frame, the $z$  axis is perpendicular to 
the decay plane of $\phi \to K^+ K^-$, and $p_y(K^+)
\geq 0$. The coordinates $(\theta, \varphi)$ describe the decay
direction of $l^+$ in the $J/ \psi$ rest frame and $\psi$ is 
the angle made by $\vec p(K^+)$ with the $x$ axis in the $\phi$ 
rest frame. With this convention,
\barr
{\bf x} = {\bf p}_{\phi} , &
{\bf y} = \frac{ {\bf p}_{K^+} - {\bf p}_{\phi} ( {\bf p}_{\phi}
                                \cdot {\bf p}_{K^+} ) }
        { | {\bf p}_{K^+} - {\bf p}_{\phi} ( {\bf p}_{\phi}
                                \cdot {\bf p}_{K^+} ) | } , &
{\bf z} = {\bf x} \times {\bf y} \nonumber \\
\sin \theta \;\; \cos \varphi =   {\bf p}_{\ell^+} \cdot  {\bf x}, &
\sin \theta \;\; \sin \varphi = {\bf p}_{\ell^+} \cdot  {\bf y}, &
\cos \theta =  {\bf p}_{\ell^+} \cdot {\bf z}~~.
\label{angdef}
\earr
Here, the bold-face characters represent {\it  unit} 3-vectors and
everything is measured in the rest frame of $J/\psi$. Also
\beq
\cos \psi = - {\bf p}'_{K^+}
                        \cdot {\bf p}'_{J/\psi},
\label{psidef}
\eeq
where the primed quantities are {\it  unit vectors}
measured in the rest frame of $\phi$.

The time dependence of the right-hand side of Eq.~(\ref{triple}) 
can be read off from Table~\ref{tab1}, where $\Delta m\equiv 
m_H-m_L>0$ is the mass difference of the mass eigenstates 
$B_s^H$ and $B_s^L$ of the $B_s$ system and  
$\overline{\Gamma}\equiv(\Gamma_H+\Gamma_L)/2$
denotes their average decay width. The phases
$\delta_1 \equiv \mbox{Arg}(A_{\|}(0)^*A_{\perp}(0))$ and
$\delta_2 \equiv\mbox{Arg}(A_0(0)^*A_{\perp}(0))$ are {\it
CP-conserving} strong phases. In the absence of final-state 
interactions -- probably not a justifiable assumption for 
$B_s \to J/ \psi \phi$ -- they are expected to be
$0\mbox{ (mod }\pi)$.

\begin{table}[t]
\begin{center}
\begin{tabular}{|c|l|}
\hline
Observable & Time evolution \\
\hline
$|A_0(t)|^2$  & $|A_0(0)|^2 \left[e^{-\Gamma_L t} -
e^{-\overline{\Gamma}t}
\sin(\Delta m t)\delta\phi\right]$\\
$|A_{\|}(t)|^2$ &$ |A_{\|}(0)|^2 \left[e^{-\Gamma_L t} -
e^{-\overline{\Gamma}t}\sin(\Delta m t)\delta\phi\right]$\\
$|A_{\perp}(t)|^2$ & $|A_{\perp}(0)|^2 \left[e^{-\Gamma_H t} +
e^{-\overline{\Gamma}t}\sin(\Delta m t)\delta\phi\right]$\\
\hline
Re$(A_0^*(t) A_{\|}(t))$ &  $|A_0(0)||A_{\|}(0)|\cos(\delta_2 -
\delta_1)\left[e^{-\Gamma_L t} - e^{-\overline{\Gamma}t}
\sin(\Delta m t)\delta\phi\right]$\\
Im$(A_{\|}^*(t)A_{\perp}(t))$ & $|A_{\|}(0)||A_{\perp}(0)|\left[
e^{-\overline{\Gamma}t}\sin (\delta_1-\Delta m t)+\frac{1}{2}\left(
e^{-\Gamma_H t}-e^{-\Gamma_L
t}\right)\cos(\delta_1)\delta\phi\right]$\\
Im$(A_0^*(t)A_{\perp}(t))$ & $|A_0(0)||A_{\perp}(0)|\left[
e^{-\overline{\Gamma}t}\sin (\delta_2-\Delta m t)+\frac{1}{2}\left(
e^{-\Gamma_H t}-e^{-\Gamma_L
t}\right)\cos(\delta_2)\delta\phi\right]$\\
\hline
\end{tabular}
\end{center}
\caption{Time evolution of the decay $B_s\to J/\psi(\to l^+l^-)
\phi(\to K^+K^-)$ of an initially (i.e.\ at $t=0$) pure $B_s$ meson.}
\label{tab1}
\end{table}

\begin{table}
\begin{center}
\begin{tabular}{|c|l|}
\hline
Observable & Time evolution \\
\hline
$|\overline{A}_0(t)|^2$  & $|A_0(0)|^2 \left[e^{-\Gamma_L t} +
e^{-\overline{\Gamma}t}\sin(\Delta m t)\delta\phi\right]$\\
$|\overline{A}_{\|}(t)|^2$ &$ |A_{\|}(0)|^2 \left[e^{-\Gamma_L t} +
e^{-\overline{\Gamma}t}\sin(\Delta m t)\delta\phi\right]$\\
$|\overline{A}_{\perp}(t)|^2$ & $|A_{\perp}(0)|^2 \left[e^{-\Gamma_H
t} -
e^{-\overline{\Gamma}t}\sin(\Delta m t)\delta\phi\right]$\\
\hline
Re$(\overline{A}_0^*(t)\overline{A}_{\|}(t))$ &
$|A_0(0)||A_{\|}(0)|\cos(\delta_2 -
\delta_1)\left[e^{-\Gamma_L t} + e^{-\overline{\Gamma}t}
\sin(\Delta m t)\delta\phi\right]$\\
Im$(\overline{A}_{\|}^*(t)\overline{A}_{\perp}(t))$ &
$-|A_{\|}(0)||A_{\perp}(0)|\left[
e^{-\overline{\Gamma}t}\sin (\delta_1-\Delta m t)-\frac{1}{2}\left(
e^{-\Gamma_H t}-e^{-\Gamma_L
t}\right)\cos(\delta_1)\delta\phi\right]$\\
Im$(\overline{A}_0^*(t)\overline{A}_{\perp}(t))$ &
$-|A_0(0)||A_{\perp}(0)|\left[
e^{-\overline{\Gamma}t}\sin (\delta_2-\Delta m t)-\frac{1}{2}\left(
e^{-\Gamma_H t}-e^{-\Gamma_L
t}\right)\cos(\delta_2)\delta\phi\right]$\\
\hline
\end{tabular}
\end{center}
\caption{Time evolution of the decay $\overline{B_s}\to J/\psi
(\to l^+ l^-)\phi(\to K^+ K^-)$
of an initially (i.e.\ at $t=0$) pure $\overline{B_s}$ meson.}
\label{tab2}
\end{table}

On the other hand, the quantity 
$\delta\phi=\phi_{\mbox{{\scriptsize CKM}}}$ (see Eq.~(\ref{phickm}))  
is a {\it  CP-violating} weak phase, which is introduced through 
interference effects between $B_s$--$\overline{B_s}$ mixing and 
decay processes. It can be expressed in terms of elements of the 
CKM matrix \cite{km,cptaggedbs} as
\beq\label{deltaphi}
\exp(i\delta\phi)=\frac{V_{ts}V_{tb}^\ast}{V_{ts}^\ast V_{tb}}
\frac{V_{cs}^\ast V_{cb}}{V_{cs}V_{cb}^\ast}
\eeq
and is very small, as can be seen easily by applying the 
Wolfenstein expansion of the CKM matrix \cite{wolf}. At leading 
order in this expansion $\delta\phi$ {\it vanishes}.
However, taking into account higher-order terms (for a treatment 
of such terms, see e.g.\ Ref.~\cite{blo}) gives a non-vanishing 
result \cite{dsnowmass93,dthesis}:
\beq\label{delphiwolf}
\delta\phi=2\lambda^2\eta={\cal O}(0.03).
\eeq
Consequently $\delta\phi$ measures simply the CKM parameter $\eta$.
Note that $\lambda=\sin\theta_{\mbox{{\scriptsize C}}}=0.22$ is 
related to the Cabibbo angle. Useful expressions for $\delta\phi$ 
can be found in Ref.~\cite{dsnowmass93},
where the following relation has been derived:
\beq
\delta\phi=2\lambda^2 R_b\sin\gamma.
\eeq
Here $\gamma$ is one angle of the ``usual'' unitarity
triangle \cite{ut}. Consequently, if the CKM-parameter $R_b$
(defined by Eq.~(\ref{rb})) is used as an input, $\delta\phi$
allows a determination of $\gamma$. That input allows, however, also
the determination of $\eta$ (or $\gamma$) from the mixing-induced
CP asymmetry of $B_d\to J/\psi K_S$ measuring $\sin2\beta$, where
$\beta$ denotes another angle of the unitarity triangle \cite{ut}. 
If one compares these two results for $\eta$ (or $\gamma$) 
obtained from $B_s$ and $B_d$ modes, respectively, a test of whether the 
$B_s$--$\overline{B_s}$ and $B_d$--$\overline{B_d}$ mixing phases
are described by the Standard Model, or receive additional
contributions from physics beyond that model can be performed. Needless to
note, a measurement of a value of $\delta\phi$ much larger than 
the Standard Model expectation of ${\cal O}(0.03)$ would anyway be a striking 
signal for new physics in $B_s$--$\overline{B_s}$ mixing.

An interesting interpretation of $\delta\phi$ has been given in
Ref.~\cite{akl}. There it was shown that $\delta\phi$ is related
to one angle in a rather squashed (and therefore ``unpopular'')
unitarity triangle. Note that terms of ${\cal O}(\delta\phi^2)$ have
been neglected in Table~\ref{tab1}.

The angular distribution for an initially present $\overline{B_s}$
meson is given by
$$
\frac{d^3 \Gamma [\overline{B_s}(t) \to J/\psi (\to l^+ l^-)
\phi (\to K^+ K^-)]}
{d \cos \theta~d \varphi~d \cos \psi}
\propto \frac{9}{32 \pi} \Bigl[~2 |\overline{A}_0(t)|^2 \cos^2 \psi
(1 -
\sin^2 \theta \cos^2 \varphi)
\hfill{ }
$$
$$
+ \sin^2 \psi \{ |\overline{A}_\parallel(t)|^2  (1 - \sin^2 \theta
\sin^2 \varphi)
+ |\overline{A}_\perp(t)|^2 \sin^2 \theta - \mbox{ Im }
(\overline{A}_\parallel^*(t) \overline{A}_\perp(t))
\sin 2 \theta \sin \varphi \}~~~
$$
\beq \label{triplecp}
+\frac{1}{\sqrt{2}}\sin 2 \psi \{\mbox{ Re }(\overline{A}_0^*(t)
\overline{A}_\parallel(t)) \sin^2 \theta
 \sin 2 \varphi + \mbox{ Im }(\overline{A}_0^*(t)
\overline{A}_\perp(t)) \sin
2 \theta \cos \varphi~\}~\Bigr]~~,
\eeq
where the angles are again defined by 
Eqs.~(\ref{angdef}) and (\ref{psidef}). The time dependence of this 
rate can be obtained 
easily with the help of Table~\ref{tab2}, where terms of 
${\cal O}(\delta\phi^2)$ have been neglected, as in 
Table~\ref{tab1}. In calculating Tables~\ref{tab1} and 
\ref{tab2} we have used the fact that 
$B_s\to J/\psi\,\phi$ (and $\overline{B_s}\to J/\psi\,\phi$) is
dominated to {\it  excellent accuracy} by a single weak amplitude, as
we have seen in Section~\ref{tme}.
Therefore we have to deal only with {\it  mixing-induced}
CP violation and there is no {\it  direct} CP violation, i.e.
$|\overline A_0(0)|=|A_0(0)|$, $|\overline A_\parallel(0)|
=|A_\parallel(0)|$ and $|\overline A_\bot(0)| =|A_\bot(0)|$.

It is important to note that the mass difference $\Delta m$ can be extracted 
from time-dependent analyses of tagged $B_s\to J/\psi\,\phi$ data 
samples~\cite{ddlr}. Previous experimental feasibility investigations for 
the extraction of $\Delta m$ focused entirely on tagged flavour-specific 
modes of $B_s$ mesons~\cite{snowmass93,lepbs}.

\subsection{Untagged decays }\label{untgd}

Combining Tables~\ref{tab1} and \ref{tab2}, we find that
the time evolution of the {\it  untagged} data sample for $f= J/\psi
(\rightarrow l^+ l^- )\phi (\rightarrow K^+ K^- )$ is given by
$$
\frac{d^3 \Gamma [f(t)]}
{d \cos \theta~d \varphi~d \cos \psi}
\propto \frac{9}{16 \pi} \Biggl[~2 |A_0(0)|^2 e^{-\Gamma_L t}\cos^2
\psi
(1 - \sin^2 \theta\cos^2 \varphi)
\hfill{ }
$$
$$
+ \sin^2 \psi \{ |A_\parallel(0)|^2  e^{-\Gamma_L t}
(1 - \sin^2 \theta \sin^2 \varphi)
+ |A_\perp(0)|^2 e^{-\Gamma_H t}\sin^2 \theta \} ~~~
$$
$$
+\frac{1}{\sqrt{2}}  \sin 2 \psi\left \{ |A_0(0)|| A_\parallel(0)|
\cos(\delta_2-\delta_1)e^{-\Gamma_L t}
\sin^2
\theta  \sin 2 \varphi \right\}
$$
$$
+\biggl\{\frac{1}{\sqrt{2}}|A_0(0)|| A_\perp(0)|\cos\delta_2\sin2\psi
\sin2\theta\cos\varphi
$$
\beq\label{tripleuntagged}
-|A_\parallel(0)|| A_\perp(0)|\cos\delta_1\sin^2\psi
\sin2\theta\sin\varphi\biggr\}\frac{1}{2}\left(e^{-\Gamma_H t}-
e^{-\Gamma_L t}\right)\delta\phi~\Biggr]~~.
\eeq
Remarkably the time dependence of the untagged rate does not depend  
on the mass difference $\Delta m$. This feature has been discussed 
within a more general framework in Ref.~\cite{bsbsbar}. 
Consequently, whereas $\Gamma_L$ and $\Gamma_H$ can be 
determined from the untagged data sample, the extraction
of $(\Delta m)_{B_s}$ requires tagging.
As has already been pointed out in \cite{fd1}, because of the lifetime
difference $(\Delta\Gamma)_{B_s}$, the untagged
decay rate [Eq.~(\ref{tripleuntagged})]
develops an interesting contribution for $t>0$,
which is proportional to the CP-violating weak phase $\delta\phi$.
It  originates from the imaginary parts of the interference terms
between $A_\perp(t)$ $(\overline{A}_\perp(t))$ and  
$A^\ast_\parallel(t)$ $(\overline{A}^\ast_\parallel(t))$, 
$A^\ast_0(t)$ $(\overline{A}^\ast_0(t))$. If 
$\Delta\Gamma\equiv\Gamma_H-\Gamma_L$
is in fact sizeable, we are optimistic that it will be possible  
to measure this effect.

\subsection{A closer look at the one-angle distribution}
The full three-angle distributions for tagged and untagged
$B_s \to J/\psi (\to l^+ l^-) \phi (\to K^+ K^-)$ decays discussed in the
previous subsections are quite complicated. A much simpler case arises
if we integrate out the two decay angles $\varphi$ and $\psi$ in 
(\ref{triple}), leading to the following {\it one-angle} distribution 
\cite{ddlr}:
\beq\label{one-angle}
\frac{d \Gamma (t)}{d \cos \theta} \propto
   (|A_0(t)|^2 + |A_{\|}(t)|^2)\frac{3}{8}(1 + \cos ^2 \theta)
                + |A_{\perp}(t)|^2 ~\frac{3}{4} \sin^2 \theta \,.
\label{single}
\eeq
Let us first briefly illustrate the angular moment analysis outlined in 
Section~\ref{ama} for this transparent one-angle distribution. In this case, 
we have 
\begin{equation}
g^{(1)}(\theta) = \frac{3}{8}\left(1 +\cos^2 \theta\right),\quad
g^{(2)}(\theta) = \frac{3}{4}\sin^2  
\theta. 
\end{equation}
Consequently, if we choose 
\begin{equation}
w^{(1)}(\theta) = 5 \cos^2 \theta -1\quad\mbox{and}\quad 
w^{(2)}(\theta) = 2 - 5\cos^2 \theta,
\end{equation}
the orthogonality relation
\beq
\int_{-1}^{+1} d(\cos \theta) w^{(i)}(\theta)
g^{(k)}(\theta) = \delta_{ik}
\eeq
is satisfied, and we obtain immediately
\barr
\label{cpeven}
|A_0(t_j)|^2 + |A_{\|}(t_j)|^2 & \propto & \sum_i
(5 \cos^2 \theta_i - 1) \\
\label{cpodd}
|A_{\perp}(t_j)|^2 & \propto & \sum_i
(2 - 5 \cos^2 \theta_i ) ~,
\earr
where the summation is over all the events in the same time bin as
$t_j$.

In the case of the {\it untagged} one-angle distribution,
the $\Delta mt$ oscillations proportional to the CP-violating weak phase 
$\delta\phi$ cancel, and the terms (\ref{cpeven}) and (\ref{cpodd}) evolve 
like $(|A_0(0)|^2 + |A_{\|}(0)|^2) e^{-\Gamma_L t}$ and 
$|A_{\perp}(0)|^2 e^{-\Gamma_H t}$,
respectively. A fit (now with only one parameter in each
time evolution) gives the decay widths $\Gamma_L$ and $\Gamma_H$ of the 
CP-even and CP-odd $B_s$ mass eigenstates, as well as the CP-even and CP-odd 
rates $|A_0(0)|^2 + |A_{\|}(0)|^2$ and $|A_{\perp}(0)|^2$, respectively. For 
limited statistics, one may want to use time moments~\cite{saunak}
\begin{equation}
f^{(n)} = \int_0^{\infty} dt~t^n~f(t)\,.
\end{equation}
The {\it  weighting-functions method} is thus an alternative to the 
{\it  two-bin} method suggested in \cite{dqstl, ddlr}. Note that we do 
not need any {\it a priori} information about the relative magnitudes of 
CP-even and CP-odd amplitudes.

In the case of {\it tagged} measurements, the integrated decay rates
\begin{equation}
\Gamma(t)=\int_{-1}^{+1}d(\cos \theta)\,\frac{d \Gamma (t)}{d \cos \theta}
\end{equation}
evolve in time for intitially present $B_s$ and $\overline{B_s}$ mesons as
\begin{eqnarray}
\Gamma(t)&\propto&\left(|A_0(0)|^2 + |A_{\|}(0)|^2\right)e^{-\Gamma_L t}+
|A_{\perp}(0)|^2e^{-\Gamma_H t}\nonumber\\
&&-\left(|A_0(0)|^2 + |A_{\|}(0)|^2-|A_{\perp}(0)|^2\right)
e^{-\overline{\Gamma}t}\sin(\Delta m t)\,\delta\phi\label{rateB}
\end{eqnarray}
and
\begin{eqnarray}
\overline{\Gamma}(t)&\propto&\left(|A_0(0)|^2 + |A_{\|}(0)|^2\right)
e^{-\Gamma_L t}+|A_{\perp}(0)|^2e^{-\Gamma_H t}\nonumber\\
&&+\left(|A_0(0)|^2 + |A_{\|}(0)|^2-|A_{\perp}(0)|^2\right)
e^{-\overline{\Gamma}t}\sin(\Delta m t)\,\delta\phi\,,\label{rateBbar}
\end{eqnarray}
respectively, where we have used Tables~\ref{tab1} and \ref{tab2}. 
Consequently, the time-dependent CP asymmetry arising in the decay 
$B_s\to J/\psi\,\phi$ takes the following form:
\begin{eqnarray}
\lefteqn{a_{\rm CP}(B_s(t)\to J/\psi\,\phi)\equiv\frac{\Gamma(t)-
\overline{\Gamma}(t)}{\Gamma(t)+\overline{\Gamma}(t)}}\nonumber\\
&&=-\,\frac{|A_0(0)|^2 + |A_{\|}(0)|^2 - |A_{\perp}(0)|^2}{\left(|A_0(0)|^2 + 
|A_{\|}(0)|^2\right)e^{-\Gamma_L t}+|A_{\perp}(0)|^2e^{-\Gamma_H t}}\,
e^{-\overline{\Gamma}t}\sin(\Delta m t)\,\delta\phi\,.\label{CP-asym}
\end{eqnarray}
Using the quantitative estimates collected in Table~\ref{predictions}, we
obtain 
\begin{equation}
\frac{|A_{\perp}(0)|^2}{|A_0(0)|^2 + |A_{\|}(0)|^2}=
0.1\,\ldots\,0.5\,. 
\end{equation}
Although these estimates suffer from large hadronic uncertainties,
they indicate that it may not be justified to neglect the CP-odd
contributions proportional to $|A_{\perp}(0)|^2$ in the time-dependent 
CP asymmetry (\ref{CP-asym}). 

The coefficient of $\sin(\Delta m t)\,\delta\phi$ in (\ref{CP-asym}) can be 
experimentally determined [for instance, from the untagged studies outlined 
above]. Thus the fundamental weak phase $\delta\phi$ can be cleanly 
extracted once the $\Delta m t$ oscillations are resolved. Future
experiments at the Tevatron and the LHC should be able to
achieve this goal. Once the $\Delta m t$ oscillations are traced,
one can alternatively perform 
a tagged, one-angle, time-dependent study to separate the CP-even and 
CP-odd contributions, from each of which $\delta\phi$ can be directly 
extracted.  The efficient 
extraction of the various observables depends on the detector configuration, 
so that other possible variations should be considered. The full angular 
distributions contain, of course, all the available information, and will 
be determined eventually. 

In order to determine $\delta\phi$ from {\it untagged} $B_s\to J/\psi\,\phi$ 
decays, where the $\Delta m t$ oscillations cancel, the observables 
corresponding to the interference terms 
$\mbox{ Im }(A^*_\parallel(t) A_\bot(t) )$ and 
$\mbox{ Im } (A^*_0(t) A_\bot(t) )$ must be studied. Valuable information 
about CP-conserving strong phases can also be obtained, thereby 
sheding light on the hadronization dynamics of $B_s\to J/\psi\,\phi$ and the
issue of ``factorization'', which predicts trivial strong phases. A set of 
weighting functions applicable to this case is given in Table~\ref{tab3}. 

\begin{table}
\begin{center}
\begin{tabular}{|c|c|}
\hline
Observables: $b^{(i)}(t)$ & $w^{(i)}(\theta, \varphi, \psi)$ \\
\hline
$|A_0(t)|^2$  & $\frac{1}{2}[5(\cos^2 \theta - \sin^2 \theta \cos
2\varphi)
-1]$\\
$|A_{\|}(t)|^2$ &$\frac{1}{2}[5(\cos^2 \theta + \sin^2 \theta \cos
2\varphi)
-1]$\\
$|A_{\perp}(t)|^2$ & $2 - 5 \cos^2 \theta$  \\
\hline
Re$(A_0^*(t) A_{\|}(t))$ & $\frac{5}{\sqrt{2}} \sin (2\psi)
\sin(2  \varphi)$\\
Im$(A_{\|}^*(t)A_{\perp}(t))$&$-\frac{5}{2} \sin(2  \theta) \sin
\varphi$ \\
Im$(A_0^*(t)A_{\perp}(t))$ & $\frac{25}{4 \sqrt{2}} \sin(2\psi)
\sin(2\theta) \cos\varphi$ \\
\hline
\end{tabular}
\end{center}
\caption{A set of weighting functions for extracting the
observables
$b^{(i)}(t)$ of the decays $B_s\to J/\psi(\to l^+l^-)\phi(\to K^+
K^-)$
and $B\to J/\psi(\to l^+l^-)K^{\ast}(\to\pi K)$.}
\label{tab3}
\end{table}

\subsection{The decay $B \to J/\psi(\to l^+ l^-) K^{\ast}(\to \pi
K)$}\label{BpsiK}

The angular distribution for $B \to J/\psi (\to l^+ l^-)
K^{*} (\to \pi K)$ takes the same form as Eq.~(\ref{triple}) if we  
use the decay angles specified in 
Eqs.~(\ref{angdef}) and (\ref{psidef})
with $\phi$ replaced by $K^{*}$ and $K^+$ replaced by the strange 
meson.

Using the same angles for $\overline{B}
\to J/\psi (\to l^+ l^-) \overline{K}^{*} (\to \pi
\overline{K})$, we obtain the analogous angular distribution to
the $\overline{B_s} \to J/\psi (\to l^+ l^-) \phi (\to K^+ K^-)$
case given in Eq.~(\ref{triplecp}).  The same weighting functions
(see Table \ref{tab3}) can therefore be used to determine the
corresponding observables in those decays. The comparison of the 
observables in these two modes would give us an idea of the 
extent of $SU(3)$ breaking.

\begin{table}[t]
\begin{center}
\begin{tabular}{|c|l|}
\hline
Observable & Time evolution \\
\hline
$|A_0(t)|^2$  & $|A_0(0)|^2 e^{-\Gamma
t}\left[1+\sin(2\beta)\sin(\Delta m t)
\right]$\\
$|A_{\|}(t)|^2$ &$ |A_{\|}(0)|^2 e^{-\Gamma t}\left[1+\sin(2\beta)
\sin(\Delta m t)\right]$\\
$|A_{\perp}(t)|^2$ & $|A_{\perp}(0)|^2 e^{-\Gamma t}\left[1-
\sin(2\beta)\sin(\Delta m t)\right]$\\
\hline
Re$(A_0^*(t) A_{\|}(t))$ &  $|A_0(0)||A_{\|}(0)|\cos(\delta_2 -
\delta_1)e^{-\Gamma t}\left[1+\sin(2\beta)\sin(\Delta m t)
\right]$\\
Im$(A_{\|}^*(t)A_{\perp}(t))$ & $|A_{\|}(0)||A_{\perp}(0)|e^{-\Gamma
t}\left[
\sin (\delta_1)\cos(\Delta m t)-\cos(2\beta)\cos(\delta_1)\sin(\Delta
m t)
\right]$\\
Im$(A_0^*(t)A_{\perp}(t))$ & $|A_0(0)||A_{\perp}(0)|e^{-\Gamma
t}\left[
\sin (\delta_2)\cos(\Delta m t)-\cos(2\beta)\cos(\delta_2)\sin(\Delta
m t)
\right]$\\
\hline
\end{tabular}
\end{center}
\caption{Time evolution of the decay $B_d\to J/\psi(\to
l^+l^-)K^{\ast0}(\to
\pi^0 K_S)$ of an initially (i.e.\ at $t=0$) pure $B_d$ meson.}
\label{tab4}
\end{table}

\begin{table}
\begin{center}
\begin{tabular}{|c|l|}
\hline
Observable & Time evolution \\
\hline
$|\overline{A}_0(t)|^2$  & $|A_0(0)|^2 e^{-\Gamma t}\left[1-
\sin(2\beta)\sin(\Delta m t)\right]$\\
$|\overline{A}_{\|}(t)|^2$ &$ |A_{\|}(0)|^2 e^{-\Gamma t}\left[1-
\sin(2\beta)\sin(\Delta m t)\right]$\\
$|\overline{A}_{\perp}(t)|^2$ & $|A_{\perp}(0)|^2 e^{-\Gamma
t}\left[1+
\sin(2\beta)\sin(\Delta m t)\right]$\\
\hline
Re$(\overline{A}_0^*(t)\overline{A}_{\|}(t))$ &
$|A_0(0)||A_{\|}(0)|\cos(\delta_2 -
\delta_1)e^{-\Gamma t}\left[1-\sin(2\beta)\sin(\Delta m t)
\right]$\\
Im$(\overline{A}_{\|}^*(t)\overline{A}_{\perp}(t))$ &
$-|A_{\|}(0)||A_{\perp}(0)|e^{-\Gamma t}\left[
\sin (\delta_1)\cos(\Delta m t)-\cos(2\beta)\cos(\delta_1)\sin(\Delta
m t)
\right]$\\
Im$(\overline{A}_0^*(t)\overline{A}_{\perp}(t))$ &
$-|A_0(0)||A_{\perp}(0)|e^{-\Gamma t}\left[
\sin (\delta_2)\cos(\Delta m t)-\cos(2\beta)\cos(\delta_2)\sin(\Delta
m t)
\right]$\\
\hline
\end{tabular}
\end{center}
\caption{Time evolution of the decay $\overline{B_d}\to J/\psi(\to
l^+l^-)
\overline{K}^{\ast0}(\to
\pi^0 K_S)$ of an initially (i.e.\ at $t=0$) pure $\overline{B_d}$
meson.}
\label{tab5}
\end{table}

If the $K^{\ast0}$ is observed to decay to the CP eigenstate
$\pi^0 K_S$, the time evolution of the corresponding three-angle
distributions [Eqs.~(\ref{triple}) and ~(\ref{triplecp})]
is given in Tables~\ref{tab4} and \ref{tab5}, 
respectively~\cite{dqstl}.
Tables~\ref{tab4} and \ref{tab5} assume that the unmixed amplitudes depend 
on a single, unique weak phase, which is justified within the CKM model 
(see Section II). In these tables, $\Gamma$ and $\Delta m>0$ describe 
$B_d$--$\overline{B_d}$ mixing. They are related to each other through 
the mixing parameter $x_d\equiv\left(\Delta m/\Gamma\right)_{B_d}$.
In analogy to Eq.~(\ref{deltaphi}), mixing-induced CP violation in
$B_d\to J/\psi(\to l^+l^-)K^{\ast0}(\to\pi^0K_S)$
 \cite{dqstl,kayser} measures a weak
phase $\tilde\beta$, which is given by
\beq\label{beta}
\exp(-2i\tilde\beta)=\frac{V_{td}V_{tb}^\ast}{V_{td}^\ast V_{tb}}
\frac{V_{cs}^\ast V_{cb}}{V_{cs}V_{cb}^\ast}.
\eeq
Within the Wolfenstein expansion \cite{wolf}, $\tilde\beta$ is equal,
to a very good approximation, to the angle $\beta$ of the
``standard'' (non-squashed) unitarity triangle \cite{ut}. 
Therefore we have not distinguished between $\tilde\beta$ and 
$\beta$ in Tables \ref{tab4} and \ref{tab5}.

Whereas the rates for tagged $B_d\to J/\psi K_S$ and
$\overline{B_d}\to J/\psi K_S$ events, which are given by
\begin{eqnarray}
\Gamma[B_d(t)\to J/\psi K_S]&\propto&|A(0)|^2 e^{-\Gamma t}
\left[1-\sin(2\beta)\sin(\Delta mt)\right]\\
\Gamma[\overline{B_d}(t)\to J/\psi K_S]&\propto&|A(0)|^2 
e^{-\Gamma t}
\left[1+\sin(2\beta)\sin(\Delta mt)\right],
\end{eqnarray}
allow {\it  only} the determination of $\sin(2\beta)$
and of $(\Delta m,\, \Gamma)_{B_d}$, an analysis of the tagged
three-angle distribution for the decay $B_d\to J/\psi(\to
l^+l^-)K^{\ast0}(\to\pi^0K_S)$ (and its CP-conjugate) yields 
valuable additional information from the interference terms, as 
can be seen by looking at Tables~\ref{tab4} and \ref{tab5}:
\begin{itemize}
\item Re$(A^\ast_0(t)A_\parallel(t))$ provides
additional information on $\cos(\delta_2-\delta_1)$.
\item Im$(A^\ast_\parallel(t)A_\perp(t))$,
Im$(A^\ast_0(t)A_\perp(t))$ provide additional information both 
on $\sin\delta_{1(2)}$ and $\cos\delta_{1(2)}$ and on $\cos(2\beta)$.
The latter quantity plays an important role to resolve discrete 
ambiguities in the determination of the CKM angle $\beta$ \cite{ddf2}.
\end{itemize}
Predictions for these observables are given in
Table~\ref{predictions}.

The largest data sample for $B_d\to J/\psi K^{\ast0}$ is, however,
not for $K^{\ast0}\to\pi^0K_S$, but for $K^{\ast0}\to\pi^-K^+$.
The complete angular distributions and time dependences for the relevant 
decay modes are given in Appendix A. 
For charged $B$ decays, the corresponding time and angular
distribution is obtained by going to the isospin - related 
mode and setting $\Delta m = 0$.
Experimental studies 
of these decays are very important, since they probe 
$\sin(\delta_{1(2)})$ and non-factorizable terms through the 
observables corresponding to the left-hand sides of 
Eqs.~(\ref{factest1})--(\ref{factest0par}) 
\cite{kornerg,yamamoto,browderh}.
The relevant information about $\delta_1$ and $\delta_2$ 
extracted from these $B$
data samples, ``tagged'' at the time of decay, can be related to 
$B_s\to J/\psi\,\phi$ by using $SU(3)$
flavour symmetry of strong interactions, and allows a determination
of $\delta\phi$ from the time evolution of even the {\it  untagged} rate
given by Eq.~(\ref{tripleuntagged}) \cite{fd1}. This approach does 
not involve the assumption of factorization, just $SU(3)$ symmetry
arguments. Unfortunately the corresponding $SU(3)$-breaking
corrections cannot be treated in a quantitative way at present.

\section{The angular distribution of the colour-allowed decays
        $B_s\to D^{\ast+}_s D^{\ast-}_s$ and
        $B\to D^{\ast+}_s \overline{D}^{\ast}$}\label{cad}

The decay of $D^{\ast \pm}_s$ is predominantly electromagnetic, i.e.\
$D^{\ast \pm}_s \to D^{\pm}_s \gamma$, whereas
$\overline{D}^{\ast}$ decays also strongly to $\overline{D} \pi$.
Therefore the  angular distributions of the
two decay modes discussed in this section are quite different from
each other. The first step in the decay chain is, however, still 
of the form $P \to V_1 V_2$ (as in $B_s \to J/\psi \phi$) and 
consequently the terms $A_0, A_{\parallel}, A_{\perp}$ retain 
the same meanings as in Eq.~(\ref{ampl}) and the same physical 
significance as in Sect.~\ref{csd}.

\subsection{The decay $B_s\to D^{\ast+}_s (\to D^+_s \gamma)
                D^{\ast-}_s (\to D^-_s \gamma)$}

Applying the same convention as in Ref.~\cite{kutschke}, we
define the coordinate system as follows: in the rest frame 
of the decaying $B_s$ meson, let the directions of
motion of $D^{\ast+}_s$ and $D^{\ast-}_s$ be $z'$ and $z''$,
respectively. In the plane transverse to $z'$ (or $z''$), choose
any direction as $y'$ and $y''$. The directions of $x'$ and $x''$
are then specified uniquely via $x' = y' \times z'$ and $x'' =
y'' \times z''$. Thus, $x'$ and $x''$ point in opposite directions.
Then $(\theta',\varphi')$ is the direction of $D^+_s$ in the rest
frame of $ D^{\ast+}_s$ in the $(x'-y'-z')$ coordinate system,
whereas $(\theta'',\varphi'')$ is the direction of $D^-_s$ in the
rest frame of $ D^{\ast-}_s$ in the $(x''-y''-z'')$ coordinate  
system.

Since the choice of directions of $y'$ and $y''$ was completely  
arbitrary,
only the combination $\chi = \varphi' + \varphi''$ of $\varphi'$ and
$\varphi''$ is physical and these two angles will appear in the
angular distribution only through $\chi$. In terms of the momenta of
particles, the angles $\theta',\theta''$ and $\chi$ can be defined  
as:
\barr
\cos \theta' = {\bf p'}_{D^+_s} \cdot {\bf p}_{D^{\ast+}_s} & , &
\cos \theta'' = {\bf p''}_{D^-_s} \cdot {\bf p}_{D^{\ast-}_s}  
\nonumber \\
\sin \theta' \sin \theta''\cos \chi & = &
        - \cos\theta'\cos\theta''
        -{\bf p'}_{D^+_s} \cdot {\bf p''}_{D^-_s} \nonumber \\
\sin \theta' \sin \theta''\sin\chi & = & ({\bf p'}_{D^+_s} \times
                        {\bf p''}_{D^-_s}) \cdot  {\bf  
p}_{D^{\ast+}_s}.
\earr
The bold-faced quantities are unit three-vectors. The unprimed quantities
are measured in the rest frame of $B_s$, single-primed quantities
in the rest frame of $D^{\ast+}_s$, and double-primed quantities in  
the rest frame of $D^{\ast-}_s$.

In terms of these angles, the angular distribution takes the form
$$
\frac{d^3 \Gamma}{d\cos \theta'~d\cos \theta''~d\chi}  \propto
        \frac{9}{64 \pi}
\left\{ 2 |A_0|^2 \sin^2 \theta'~ \sin^2 \theta'' \right.
$$
$$       + \frac{1}{2}|\apar|^2 [ (1 + \cos ^2 \theta')
        (1 + \cos ^2 \theta'') +
        \sin^2 \theta'~ \sin^2 \theta''~ \cos 2\chi ] 
$$
$$
        + \frac{1}{2}|\aperp|^2 [(1 + \cos ^2 \theta')
        (1 + \cos ^2 \theta'') -
        \sin^2 \theta'~ \sin^2 \theta''~ \cos 2\chi ]
$$
$$
        - \mbox{ Im }(\apar ^* \aperp)
                  \sin^2 \theta'~ \sin^2 \theta''~ \sin 2\chi
        +
        2\sqrt{2} \mbox{ Re }(A_0^* \apar) \sin \theta'~ \sin
        \theta'' ~\cos \theta'~ \cos \theta''~ \cos \chi
$$
\beq
\left. -\, 2\,\sqrt{2}\mbox{ Im }(A_0^* \aperp) \sin \theta'~ \sin
        \theta''~ \cos \theta'~ \cos \theta''~ \sin \chi
                \right\}  ~,
\eeq
where the time dependence of all observables is implicit. It can be
read off from Table~\ref{tab1}. The weighting functions are
listed in Table~\ref{tab:pgpg}.

\begin{table}
\begin{center}
\begin{tabular}{|l|c|}
\hline
$b^{(i)}(t)$ & $w^{(i)}(\theta',\theta'',\chi)$ \\
\hline
$|A_0|^2$ & $ (-45/92) C(\theta', \theta'') +
        (245/92) \sin ^2 \theta' \sin ^2 \theta''$ \\
$|A_{\|}|^2$ & $ (10/23)  C(\theta', \theta'') -
        (45/46) \sin^2 \theta'~ \sin^2 \theta'' +
        (25/4) \sin^2 \theta'~ \sin^2 \theta''~ \cos 2\chi $ \\
$|A_{\perp}|^2$ & $ (10/23)  C(\theta', \theta'')  -
        (45/46) \sin^2 \theta'~ \sin^2 \theta'' -
        (25/4) \sin^2 \theta'~ \sin^2 \theta''~ \cos 2\chi $ \\
\hline
Re$(A_0^*A_{\|}) $ & $25 \sqrt{2} \sin \theta'~ \sin \theta''~
        \cos \theta'~ \cos \theta''~\cos \chi$  \\
Im$(A_{\|}^*A_{\perp})$ & $ - (25/4) \sin^2 \theta'~ \sin^2 \theta''~
                \sin 2\chi $ \\
Im$(A_0^*A_{\perp})$ & $ - 25 \sqrt{2} \sin \theta'~ \sin \theta''~
        \cos \theta'~ \cos \theta''~\sin \chi $ \\
\hline
\end{tabular}
\end{center}
\caption{A set of weighting functions for $B_s\to D^{\ast+}_s (\to
D^+_s \gamma)
                D^{\ast-}_s (\to D^-_s \gamma)$. 
Here $C(\theta', \theta'') = (1 + \cos^2 \theta')(1 + \cos^2 \theta'')$.}
\label{tab:pgpg}
\end{table}

The angular distribution for the CP-conjugate process
$\overline{B_s}\to D^{\ast+}_s (\to D^+_s \gamma)
                D^{\ast-}_s (\to D^-_s \gamma)$ is given by

$$
\frac{d^3 \Gamma}{d\cos \theta'~d\cos \theta''~d\chi}  \propto
        \frac{9}{64 \pi}
\left\{ 2 |\overline{A_0}|^2 \sin^2 \theta'~ \sin^2 \theta'' \right.
$$
$$       + \frac{1}{2}|\overline{\apar}|^2 [ (1 + \cos ^2 \theta')
        (1 + \cos ^2 \theta'') +
        \sin^2 \theta'~ \sin^2 \theta''~ \cos 2\chi ]
$$
$$
        + \frac{1}{2}|\overline{\aperp}|^2 [(1 + \cos ^2 \theta')
        (1 + \cos ^2 \theta'') -
        \sin^2 \theta'~ \sin^2 \theta''~ \cos 2\chi ]
$$
$$
        - \mbox{ Im }(\overline{\apar} ^* \overline{\aperp})
                  \sin^2 \theta'~ \sin^2 \theta''~ \sin 2\chi
        +
        2\sqrt{2} \mbox{ Re }(\overline{A_0}^* \overline{\apar}) 
\sin \theta'~ \sin
        \theta'' ~\cos \theta'~ \cos \theta''~ \cos \chi
$$
\beq
\left. -\, 2\,\sqrt{2}\mbox{ Im }(\overline{A_0}^* \overline{\aperp}) 
\sin \theta'~ \sin
        \theta''~ \cos \theta'~ \cos \theta''~ \sin \chi
                \right\}  ~.
\eeq
The time evolution of the various quantities is the same as in
Table~\ref{tab2}.

As in the case of $B_s \to J/\psi \phi$, the decay
$B_s\to D_s^{\ast+}D_s^{\ast-}$ is dominated by a single weak 
amplitude. Therefore the analysis of the tagged and untagged 
decays outlined in Sect.~\ref{tgd} and \ref{untgd} remains 
valid by replacing $(J/\psi, \phi, l^+, l^-, K^+, K^-) \to
(D^{\ast+}_s, D^{\ast-}_s, D^+_s, \gamma, D^-_s, \gamma)$.
Since this process is colour-allowed, factorization is expected to  
hold more strongly.

\subsection{The decay $B \to D^{\ast+}_s (\to D^+_s \gamma)
        \overline{D}^{\ast}(\to \overline{D} \pi)$}\label{bdd}

Whereas the decay of the $D^{\ast+}_s$ meson is of the form $V \to
P \gamma$, which has the same angular dependence as $V \to l^+ l^-$  
for massless leptons, the $\overline{D}^{\ast}$ decay belongs 
to the category $V \to P_1 P_2$. The net angular distribution
should therefore have the same form as that for $P \to V(\to l^+ l^-)
V(\to P_1 P_2)$. The angular distribution is thus given by
Eq.~(\ref{triple}), where the definitions of angles are the same as
in Eqs.~(\ref{angdef}) and (\ref{psidef}) with $\phi$ replaced by
$\overline{D}^{\ast}$, $l^+$ replaced by $D^+_s$, and $K^+$ 
replaced by the charmed meson arising from the 
$\overline{D}^{\ast}$ decay. The angular distribution for the 
CP-conjugate decay $\overline{B}\to D^{\ast-}_s (\to D^-_s 
\gamma) D^{\ast}(\to D \pi)$ is given as in Eq.~(\ref{triplecp}). 
The weighting functions collected in Table~\ref{tab3} can 
be used to extract the corresponding observables from 
experimental data. 

At this point, a few comments concerning the time evolution of these angular
distributions are in order. Let us first consider decays of neutral $B_d$
mesons. Since here the final states are flavour-specific, no interference 
effects between $B_d$--$\overline{B_d}$ mixing and decay processes arise in 
this case. Consequently, the time evolution of the corresponding observables 
is only governed by the ``mixing'' of the initial particle, which is either 
a pure $B_d$ or $\overline{B_d}$. For $B_d\to D^{\ast+}_s (\to D^+_s \gamma)
D^{\ast-}(\to \overline D \pi)$ and
$\overline{B_d} \to D^{\ast-}_s (\to D^-_s \gamma){D}^{\ast+}(\to D \pi)$, 
the time evolution of a generic observable ${\cal Q}$ of the angular
distribution takes hence the form ${\cal Q}(t) = {\cal Q}(0)\,e^{-\Gamma t} 
\,\cos^2 (\Delta m t/2)$, while we have $ {\cal Q}(t) = {\cal Q}(0)\, 
e^{-\Gamma t}\, \sin^2 (\Delta m t/2)$ in the case of 
$B_d \to D^{\ast-}_s (\to D^-_s \gamma){D}^{\ast+}(\to D \pi)$
and $\overline{B_d} \to D^{\ast+}_s (\to D^+_s \gamma)
D^{\ast-}(\to \overline D \pi)$. The time evolutions of the
untagged flavour-specific decays and the related $B^\pm$ decays can be 
obtained straightforwardly from these expressions by setting $\Delta m=0$.

\newpage

\section{Summary}\label{sum}

The kinematics of $B$ and $B_s$ meson decays into two vector-particles, 
which both continue to decay through CP-conserving 
interactions into two lighter particles, involve three 
independent decay angles. The time evolution of the coefficients 
of the corresponding angular distributions contains valuable 
information about the lifetime and mass differences between 
the $B_s$ mass eigenstates $B_s^H$ and $B_s^L$, the relative 
magnitudes and phases of CP-odd and CP-even decay amplitudes,
and CP-violating effects, including the Wolfenstein parameter 
$\eta$ and the CKM angle $\beta$. The ratios of these coefficients 
are estimated by using various form-factor models.
Determinations of these time-dependent coefficients will be 
useful in testing these models and furthermore in determining 
the extent to which factorization or the $SU(3)$ flavour symmetry 
of strong interactions hold in these decays.

The observables of the angular distributions can be 
determined from experimental data by an angular-moment 
analysis in which 
the data are weighted by judiciously chosen weighting functions 
in order to arrive {\it directly} at the observables. At times, 
this permits the extraction of the fundamental CKM parameters.
A method applicable to all kinds of angular distributions is  
indicated, where the weighting functions can be determined 
without any {\it a priori} knowledge of the values of the 
coefficients. This method is almost as good
as the likelihood-fit method for a small number of parameters 
and is expected to give some reliable results even with low 
statistics where a likelihood fit to a large number of 
parameters is inefficient.

The $B_s$ meson decays $B_s \to J/\psi \phi, ~ D_s^{*+}D_s^{*-}$ 
are considered in the light of a possible width difference
$(\Delta \Gamma)_{B_s}$. The observables of their angular  
distributions can be related to those of the decays 
$B \to J/\psi K^*,D_s^{*+}\overline{D}^*$ by using the $SU(3)$ 
flavour symmetry, where $B$ stands for $B_d$ or $B^+$. The full 
angular distributions for all these transitions are given 
explicitly, and the corresponding weighting functions are 
specified. The time-dependent observables in all these
decays provide information about the corresponding values of 
$\Delta\Gamma$ and $\Delta m$. In addition, the decays of $B_s$ mesons 
inform us about the Wolfenstein parameter $\eta$, while 
$B_d \to J/\psi K^{\ast}(\to\pi^0 K_S)$ probes the CKM angle $\beta$. 
Some of the quantities related 
to the $B_s$ case can even be extracted from {\it untagged} data 
samples, where one does not distinguish between initially present 
$B_s$ or $\overline{B_s}$ mesons. The comparison between 
coefficients of angular distributions of $B_s$ and $B$ mesons may 
give us an idea about $SU(3)$-breaking effects, while the comparison 
of colour-suppressed $(B \to J/\psi V)$ and colour-allowed $(B \to 
D_s^{\ast \pm} V)$ modes should help in testing the expectation that 
factorization holds to a greater extent in the latter case.

\section* {Acknowledgements}
We are grateful to H.J.~Lipkin and J.L.~Rosner for previous collaborations.
We thank D.~Atwood, J.~Incandela, R.~Kutschke, J.~Lewis, M.~Neubert, K.~Ohl, 
S.~Pappas, M.~Schmidt, S.~Sen, M.~Shochet, W.~Wester, H.~Yamamoto for useful 
discussions. This work was supported in part by the Department of Energy, 
Contract Nos.~DE-AC02-76CHO3000 and DE FG02 90ER40560.

\section*{Appendix}

\appendix
\label{append}

\section{Angular distributions and time dependences for flavour-specific 
$B_d \to J/\psi(\to l^+ l^-)\, K^{\ast}(\to K^\pm \pi^\mp)$ modes}

The angles are defined as in (\ref{angdef}), where the $\phi$ meson 
is replaced by $\stackrel{(-)}{K^{\ast}}$, and the $K^+$ meson 
by the strange meson in the final state. In order to parametrize the 
corresponding angular distributions, we use the following combinations of
trigonometric functions: 
\barr
f_1 & = & 2\, \cos^2\psi\, (1 - \sin^2\theta \,\cos^2\varphi) 
\nonumber \\
f_2 & = & \sin^2\psi\, (1 - \sin^2\theta \,\sin^2\varphi)
\nonumber \\
f_3 & =  & \sin^2\psi \,\sin^2\theta
\nonumber \\
f_4 & = & \sin^2\psi\,\sin2\theta\, \sin\varphi
\nonumber \\
f_5 & = & (1/\sqrt{2}) \,\sin2\psi\,\sin^2\theta \sin2 \varphi
\nonumber \\
f_6 & = & (1/\sqrt{2}) \,\sin2\psi\, \sin2 \theta\, \cos\varphi\,.
\earr
Taking into account $|\overline{A_f}| = |A_f|$ and using the notation
$A_f\equiv A_f(0)$, where $f\in\{0, \|, \perp\}$, we obtain 
\barr
\lefteqn{\frac{d^3 \Gamma [B_d(t) \to J/\psi (\to l^+ l^-) K^{\ast} 
(\to K^+ \pi^-)]}
{d \cos \theta~d \varphi~d \cos \psi}=
\frac{9}{32 \pi}\, \cos^2\left(\frac{\Delta m t}{2}\right)\, e^{-\Gamma t}}\\
&&\times \left\{ f_1 |A_0|^2   + f_2 |A_{\|}|^2 + f_3 |A_{\perp}|^2 
 - f_4\, {\mbox  Im}\,(A_{\|}^{\ast} A_{\perp})  
 + f_5\, {\mbox Re}\,(A_0^{\ast} A_{\|})
 + f_6\, {\mbox Im}\,(A_0^{\ast} A_{\perp}) \right\}\nonumber
\earr         
\barr
\lefteqn{\frac{d^3 \Gamma [\overline{B}_d(t) \to J/\psi (\to l^+ l^-) 
\overline{K^{\ast}} 
(\to K^- \pi^+)]}
{d \cos \theta~d \varphi~d \cos \psi}=\frac{9}{32 \pi}\,
\cos^2\left(\frac{\Delta m t}{2}\right)\, e^{-\Gamma t}}\\
&&\times\left\{ f_1 |A_0|^2   + f_2 |A_{\|}|^2 + f_3 |A_{\perp}|^2 
 + f_4\, {\mbox  Im}\,(A_{\|}^{\ast} A_{\perp})  
 + f_5\, {\mbox Re}\,(A_0^{\ast} A_{\|})
 -  f_6\, {\mbox Im}\,(A_0^{\ast} A_{\perp}) \right\}\nonumber
\earr           
\barr
\lefteqn{\frac{d^3 \Gamma [B_d(t) \to J/\psi (\to l^+ l^-) \overline{K^{\ast}} 
(\to K^- \pi^+)]}{d \cos \theta~d \varphi~d \cos \psi}=
\frac{9}{32 \pi}\,\sin^2\left(\frac{\Delta m t}{2}\right)\,e^{-\Gamma t}}\\
&&\times \left\{ f_1 |A_0|^2  + f_2 |A_{\|}|^2 + f_3 |A_{\perp}|^2 
 + f_4\, {\mbox  Im}\,(A_{\|}^{\ast} A_{\perp})  
 + f_5\, {\mbox Re}\,(A_0^{\ast} A_{\|})
 -  f_6\, {\mbox Im}\,(A_0^{\ast} A_{\perp}) \right\}\nonumber
\earr           
\barr
\lefteqn{\frac{d^3 \Gamma [\overline{B}_d(t) \to J/\psi (\to l^+ l^-) K^{\ast} 
(\to K^+ \pi^-)]}
{d \cos \theta~d \varphi~d \cos \psi}= 
\frac{9}{32 \pi}\,\sin^2\left(\frac{\Delta m t}{2}\right)\,e^{-\Gamma t}}\\
&&\times\left\{ f_1 |A_0|^2  + f_2 |A_{\|}|^2 + f_3 |A_{\perp}|^2 
 - f_4\, {\mbox  Im}\,(A_{\|}^{\ast} A_{\perp})  
 + f_5\, {\mbox Re}\,(A_0^{\ast} A_{\|})
 + f_6\, {\mbox Im}\,(A_0^{\ast} A_{\perp})\right\}.\nonumber
\earr

\end{document}